# SURVIVING IN OCEAN WORLDS: EXPERIMENTAL CHARACTERIZATION OF FIBER OPTIC TETHERS ACROSS EUROPA-LIKE ICE FAULTS AND UNRAVELING THE SLIDING BEHAVIOR OF ICE

V. Singh[1], C. McCarthy[1], M. Silvia[2], M.V. Jakuba[2], K.L. Craft[3], A.R. Rhoden[4], C.R. German[2], and T. Koczynski[1].

[1] Corresponding author: vsingh@ldeo.columbia.edu
[1] Lamont-Doherty Earth Observatory, 61 Route 9W, Palisades, NY 10964
[2] Woods Hole Oceanographic Institution, Woods Hole, MA 02543
[3] Johns Hopkins University Applied Physics Laboratory
[4] Southwestern Research Institute – Boulder

## Abstract

As an initial step towards in-situ exploration of the interiors of Ocean Worlds to search for life using cryobot architectures, we test how various communication tethers behave under potential Europa-like stress conditions. By freezing two types of pretensioned insulated fiber optic cables inside ice blocks, we simulate tethers being refrozen in a probe's wake as it traverses through an Ocean World's ice shell. Using a cryogenic biaxial apparatus, we simulate shear motion on pre-existing faults at various velocities and temperatures. These shear tests are used to evaluate the mechanical behavior of ice, characterize the behavior of communication tethers, and explore their limitations for deployment by a melt probe. We determine (a) the maximum shear stress tethers can sustain from an ice fault, prior to failure (viable/unviable regimes for deployment) and (b) optical tether performance for communications. We find that these tethers are fairly robust across a range of temperature and velocity conditions expected on Europa (T(K) = 95 to 260; velocity (m/s) = $5 \times 10^{-7}$ to $3 \times 10^{-4}$). However, damage to the outer jackets of the tethers and stretching of inner fibers at the coldest temperatures tested both indicate a need for further tether prototype development.

Overall, these studies constrain the behavior of optical tethers for use at Ocean Worlds, improve the ability to probe thermomechanical properties of dynamic ice shells likely to be encountered by landed missions, and guide future technology development for accessing the interiors of (potentially habitable ± inhabited) Ocean Worlds.

Subject Keywords: Cryobots, Tethers, Communication, Europa, Ice Shell, Shear Stress

1. **Introduction**

Ocean worlds, like Jupiter's moon Europa, may harbor conditions conducive to life, including a global ocean of liquid water (Pappalardo et al. 1999; Hand et al. 2009 & 2022, and references therein) where nutrients may potentially be available including essential chemistry, sources of heat or energy, and time (Anderson et al. 1998; Greenberg et al. 2000; Chyba & Phillips 2001a, b). For Europa, as in many bodies, heat is generated through tidal dissipation due to the eccentric orbit about its parent planet (Ojakangas & Stevenson 1989; Soderlund et al. 2014). The orbit's eccentricity means the distance to the parent body varies, leading the moon to change its shape in response, which generates internal friction within the moon's interior that provides heat to help maintain the ocean (Greenberg et al. 1998) and causes fracturing and potential fault motion (e.g., Rhoden et al. 2012; Hammond et al. 2020; Lien et al. 2022). Predicted contact of the (salty) ocean with a rocky silicate core can potentially create a geochemically rich environment, akin to conditions that arise in association with seafloor fluid flow on Earth's oceans (German & Seyfried 2014; On Europa: Lowell & DuBose 2005), which could provide energy for life through redox cycles (e.g., Vance et al. 2016). The overlying insulating ice shell is also ideal for geologically-short exchange processes between the surface and subsurface. Hence, the exploration of Ocean Worlds are particularly intriguing to us as potential abodes of life beyond Earth. Although we focus here on Europa, there is evidence for subsurface oceans within Enceladus (in contact with a rocky seafloor) (e.g., Hsu et al. 2015; Waite et al. 2017; Patthoff et al. 2011; Postberg et al. 2009, 2018), Mimas (Rhoden & Walker 2022), Dione (Beuthe et al. 2016), Titan, Ganymede, Callisto (Nimmo & Pappalardo 2016), and potentially others including Triton and Pluto (Hussman et al. 2006; Nimmo & Spencer 2015).

Accessing the subsurface ocean or melt pockets within an ice shell on Europa (or other Ocean Worlds) poses significant challenges. From a technological perspective, a successful exploration mission will require starting in vacuum at cryogenic temperatures, penetrating 10's of km through an ice shell and diving 100's of kms through the ocean to explore the seafloor (Anderson et al. 1998), while maintaining communication. This journey entails navigating through an ice shell with unknown thermophysical and mechanical properties (Henderson et al. 2019), surviving tidal stressing, potential faulting with cm's of displacement, and other resurfacing events, and/or exposure to salts, sulfuric acid and other potentially caustic chemistries (Kargel et al. 2000; Zolotov & Shock 2001; Marion 2002; Zolotov & Kargel 2009; Vu et al. 2016). Coupled with potential plume activity (e.g., Jia et al. 2018; Roth et al. 2014), these are substantial challenges to overcome. A robust communication strategy and hardware that can provide data transmission rates adequate to achieve science and exploration objectives are critical for any future missions to access the subsurface of an Ocean World.

A robust technology concept has emerged for effectively exploring the interior of the ice shell: the ice penetrating robot, or "cryobot" (e.g., Howell et al. 2020). Concepts employ fiber optic tethers (Figure 1), which we have evaluated in this work, and/or radio frequency (RF) free-space relay modules proposed to enable communication between a descending probe and a surface lander (Bryant et al. 2002; Cwik et al. 2018; Oleson et al. 2019). A cryobot penetrates through the ice by melting, excavating icy material (and non-icy contaminants), or a hybrid method. Terrestrial ice-probes have been proposed and/or developed for decades (e.g., Philberth Probe, Aamot 1967; Cryobot, Zimmerman et al. 2001; Ice Diver, Winebrenner et al. 2013; VALKYRIE, Stone et al. 2014), with recent field work in Antarctica and Greenland exploring the viability of the concept for future planetary missions (Bar-Cohen & Zacny 2009; Zacny et al. 2016, SLUSH, 2018). However, these technologies require sustained development and integration to mature the cryobot system.

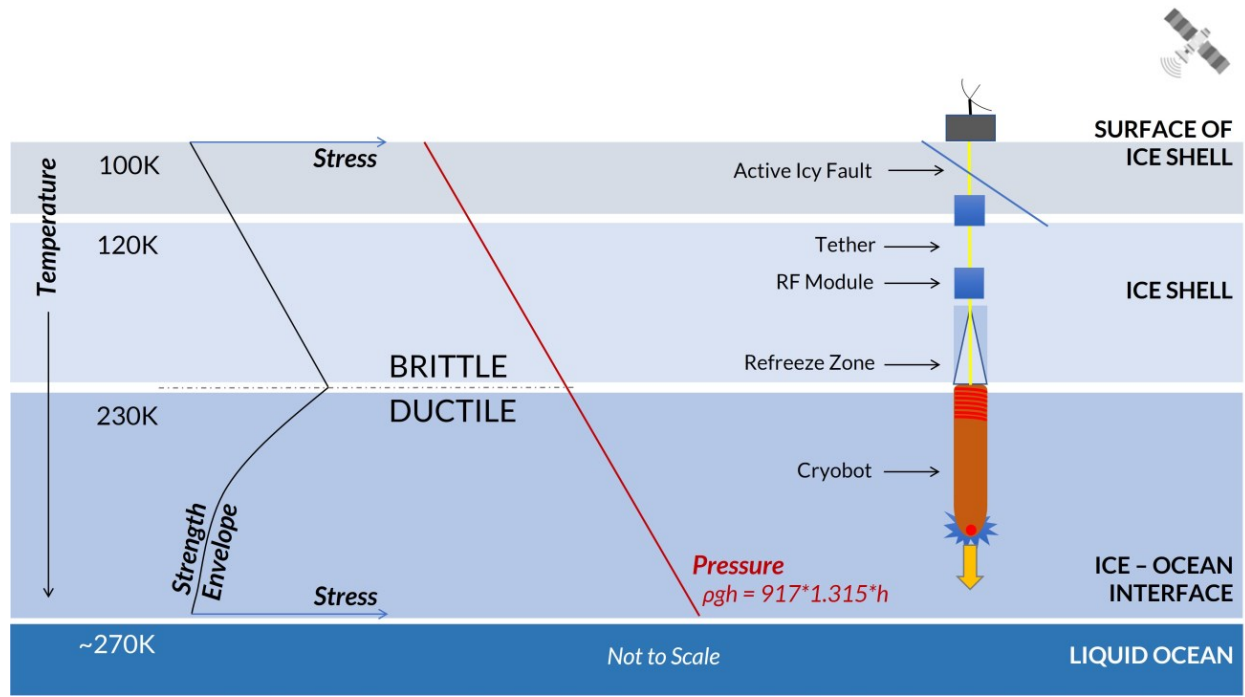

**Figure 1:** An illustration of a cryobot concept with communication tethers deployed in Europa's potentially tectonic ice shell, highlighting the challenges of surviving across an extreme temperature-pressure gradient and varying lithospheric strength of the shell (Dombard & McKinnon 2006).

Maintaining communication over the duration of an exploration campaign is a central component of mission design. In particular, tethered communication techniques offer unparalleled data transfer rates (high bandwidth communication) for minimal size, mass, and power, and have great heritage from state-of-the-art terrestrial ocean exploration. Fiber optic micro-tethers of sufficient length (20+ km) and low mass to support Europa cryobot mission concepts (e.g., Wilcox 2017; Oleson et al. 2019) have already deployed here on Earth: to reach the deepest parts of our ocean (~ 11 km); to explore from the ice-water interface to the seafloor beneath the ice-covered Arctic Ocean (Nereus - Bowen et al. 2009; NUI - Jakuba et al. 2018); and in km-deep boreholes in Greenland and Antarctica for glaciological applications of distributed strain sensing (Liehr et al. 2009), acoustic sensing and temperature sensing (Tyler et al. 2013; Kobs et al. 2014; Booth et al. 2020).

These communication tethers have survived at temperatures in the range of ~250 to 273 K, pressures up to 100 MPa, and surface ice velocities of 550 m a$^{-1}$ (1.7×10$^{-5}$ m/s), demonstrating the feasibility of using fiber optic tethers to explore (and study thermal processes in) a high deformation environment. However, their ability to withstand in-ice conditions on Europa has not yet been demonstrated, and their long-term durability remains to be established, with failure observed in a glacial field campaign in West Greenland as early as ~6 weeks (e.g., due to onset of temperature gradient vacillations as opposed to potential strain related failure) (International Thwaites Glacier Collaboration, Law et al. 2020). Other key concerns identified include signal losses in the cable, high mass/volume (i.e., meeting spaceflight mass constraints), and significant risk of tether damage due to shear stress or refreezing of the ice-channel during the mission descent phase (Dachwald et al. 2020).

To ascertain the feasibility of employing various forms of optical communication on Europa and other Ocean Worlds, our multi-disciplinary team has tested the mechanical and data transfer performance of lightly armored commercial optical tethers that have recently been introduced into the marine robotics domain (Bowen et al. 2009; Jakuba et al. 2018). The goal is to ascertain how these commercial tethers might perform across a pre-existing active fault in the brittle portion of an ice shell—identified as a mission risk scenario — and then apply the information learned to develop improved tethers and a multi-modal communication strategy.

## 2. Anticipated Conditions on Ocean Worlds like Europa and Enceladus

Conditions on Europa and other icy Ocean Worlds are considerably different from terrestrial conditions. Characterizing these ice shell environments at the surface and in the interior will provide key engineering and design constraints for ice and ocean access technologies under development and help retire associated technological risks (Howell et al. 2020). In this section, we explore the environmental conditions expected in Europa's ice shell, which motivate the design of our laboratory experiments and constrain the test parameters, as described in Section 3.

At the near-surface of the ice shell (Figure 1), the temperature is expected to lie in the range of 70-90 K at polar latitudes and up to 100-125 K at equatorial latitudes, over an Europan day, with an average surface temperature of ~100 K (Spencer et al. 1999, Rathbun et al. 2010; Berdis et al. 2021). The interior of the ice shell maintains warmer temperatures of 120-220 K that increases towards its base where, near the ice-ocean interface, temperatures reach about 230-270 K. At the ice-ocean interface, we expect the solid and liquid phases of water to co-exist, with temperatures at the applicable melting point, which is affected by both the presence of salts and the pressure. At a depth of 10 to 30 km of the ice shell below Europa's surface, a pressure of ~13 to 40 MPa is estimated, which would depress the melting point to 270-272 K, while salts such as NaCl could depress this melting point to as low as ~250 K. So, the temperature at the interface will be between 250 K and 270 K, depending on composition. These sets of conditions, cumulatively, represent the homologous temperature range (equation 1):

$$T/T_m = 0.34\text{-}0.99 \quad \ldots \text{(Equation 1)}$$

where homologous temperature is the temperature experienced by the material (T) relative to its melting temperature ($T_m$).

Therefore, like rock on the earth, the ice in the shell will transition between brittle rheology at the surface to ductile rheology at depth. The transition from brittle to ductile behavior has mostly been studied by extrapolating behavior in either regime and determining the depth at which the strengths likely intersect (Brace & Kohlstedt 1980; Dombard & McKinnon 2006). The thermal gradient of Europa is unknown and depends on whether the ice shell is purely conductive or the lower region is convecting. In the simplest case, Dombard & McKinnon (2006) assume a linear relationship and estimated a thermal gradient of 18K/km and a strain rate of $1\times10^{-15}s^{-1}$ for a pure water ice shell. The strain rate here is the estimated shortening rate for contractional folding, considered typical for orogenic processes on Earth (Pffifner & Ramsay 1982), with build-up timescales of 3 kyr to 10-60 Myr (Hoppa et al. 1999; Ojakangas & Stevenson 1989). Using laboratory values for the frictional strength of ice (i.e., Byerlee's law; Beeman et al. 1988) in the brittle regime and the viscous flow law (Goldsby & Kohlstedt 2001) for the ductile regime, they determined a lithospheric strength envelope for ice on Europa, based on its dependence on depth: increasing strength due to increasing lithostatic pressure vs decrease in strength due to temperature increase promoting ductile flow. They predicted the brittle-ductile transition (BDT) to occur at ~2.4 km depth from the surface (for an ice shell thickness of ~10.7 km, surface temperature of 80 K, grain size of 200 microns).

The mechanics of several classes of tidally stressed faults on Europa are poorly understood. For instance, cycloids are a class of features that are inferred to slip at rates of a few m/s (Hoppa et al. 1999; 2000). A qualitative concept called tidal "walking" was introduced to describe how the fault normal and fault shear stresses drive open/close and strike-slip motions (e.g., Tufts et al. 2000). Improved models of strike-slip motion on icy faults include friction and the Coulomb criterion to determine whether and when (in the tidal cycle) sliding occurs (e.g., Smith-Konter & Pappalardo 2008; Olgin et al. 2011; Rhoden et al. 2012). Additionally, a reconstruction of the northern Falga Regio provided strong evidence for a subduction-like process, which implies plate-scale thrust faulting (Kattenhorn & Prockter 2014). Subsequent studies are identifying additional evidence for plate-like motions (e.g., Collins et al. 2016). Although evidence is mounting that Europa's ice shell is a regionally active mobile lid, the proposed models cannot predict whether earthquake-like stick-slip motion can occur, nor the location with depth at which this would be expected. However, cracks are prevalent at the surface of Europa (Katterhorn & Hurford 2009), hence for a mission to its ocean, any probe and its communication hardware must be robust to the potential for this type of activity.

### 3. Materials and Methods
*3.1 Description of Tethers Selected*

To assess the robustness to shear of optical fiber tethers, we selected two types of single mode tethers from Linden Photonics Inc.: (i) Strong Tether Fiber Optic Cable (STFOC; cable ID: 1511006A / LINDEN-SPE-7034) and (ii) High Strength STFOC, a kevlar-reinforced STFOC referred to here as HS STFOC (cable ID: 1909087 / LINDEN-SPE-7052). Similar designs have been qualified for space flight (O'Riorden & Mahapatra 2012), and have previously proven effective for use under hydrostatic pressures in deep-sea, under-ice, ocean submersibles (Bowen et al. 2019; Jakuba et al. 2018). Linden is also able to supply its cables on self-supporting spools, normally for deployment from undersea vehicles, but likely adaptable to a cryobot assuming the mission architecture where payout occurs in liquid water melt behind the vehicle.

The cables tested are relatively lightweight and flexible, with standard optical propagation loss over kilometers of cable (STFOC attenuation at 1310 nm is < 0.45 dB/km and at 1550 nm is <0.35 dB/km (Linden Photonics, Inc. 2021 Cable Catalog)). The STFOC (Figure 2a) includes a tight buffered (no excess fiber length), bare fiber with a hermetic primary strength coating (jacket) of Liquid Crystal Polymer (LCP), which protects the inner fiber from moisture (e.g., reaction with hydrogen leads to transmission loss over time). The LCP is surrounded by an abrasion and chemical resistant outer layer. The HS STFOC (Figure 2b) is representative of an armored cable that includes an additional layer of Kevlar strength members braided over the top, with a final outer jacket of thermoplastic polyurethane (TPU). These tethers range in diameter (OD) from 0.76 mm for the STFOC to 1.9 mm for lightly armored HS STFOC cable, with a mass of 0.6 and 3.6 kg/km, respectively. They are capable of supporting tensile loads of 50 lbs (STFOC), up to 250 lbs (HS STFOC), as tested at ambient conditions (Linden Photonics, Inc. 2021 Cable Catalog).

Prior to this study, the shear strength of tethers had not been reported at any conditions. The optical communication performance and any potential decays arising due to long soaks in ice for depths of 20+ km over the suggested 3-year period of a cryobot mission were also not known.

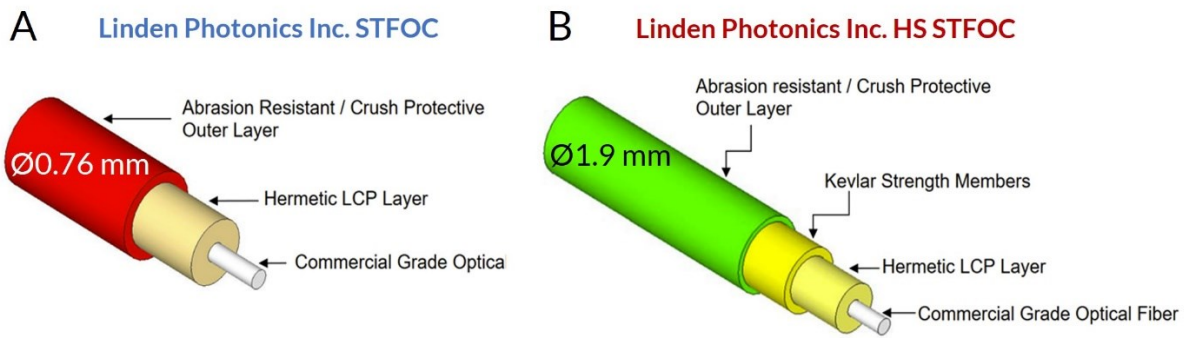

**Figure 2:** Layers of a Linden Photonics Inc. (A) Strong Tether Fiber Optical Cable (STFOC) and (B) High Strength Strong Tether Fiber Optic Cable (HS STFOC). Images modified from Linden Photonics, Inc. 2021 Cable Catalog.

*3.2 Sample Preparation*

To test the strength of tethers fully embedded in ice, we created a custom, aluminum (6061-T6) three-section die (Figure 3) that applies pretension to the tether, around which we freeze a polycrystalline ice sample with two pre-existing faults in a vertical orientation. Our design takes advantage of relevant double shear testing methods developed previously to test cables embedded in concrete (e.g., Mirzaghorbanali et al. 2017; Rasekh et al. 2017). For example, the die has a central to side ice block length ratio of 3:2 (4.95 to 3.3 cm), with a height and width of ~4.95 cm, and a total sample length of 29 cm (Figure 5). 11.55 cm of tether was embedded in these ice blocks. There are two 3.8 cm diameter mandrels at the outer ends, around which an additional ~60 cm of tether was wrapped and pre-tensioned per mandrel. The diameter was chosen based on the manufacturer's recommended turn radius of the tether. A combination of a plug, fairlead and retainer components at both ends of the ice blocks allowed the tether to run out of the mold, while maintaining vacuum inside. A strap held the tether down without crimping and the mandrels were rotated and locked into place with set screws to pre-tension the tether within the mold – ensuring no extra tether was pulled into the mold to accommodate the strain during testing. The total length of tether between the pigtail connectors was ~200 cm.

To ensure uniform ice with controlled properties, we used the "standard ice" method of ice fabrication (Cole 1979), in which we utilized ground seed ice that was sieved between 250 and 500 μm mesh sieves and lightly packed into the die (Figure 4). The mold was connected to a vacuum at the top, and a volume of deionized water was introduced from the bottom via tubing and a two-position valve: one position to flood the center section and the second position to flood the outer sections. The die and floodwater were first equilibrated at 273.15 K for ~30 minutes prior to flooding. After flooding was complete, the die was placed on a cold plate within a freezer, with insulation around the sides to promote directional freezing to remove air/impurities. The sections were frozen in steps: first the central block surrounded by vertical gates, then the outer blocks with spacers (replacing the gates) and interfaces lubricated with a release agent or oil. This die and protocol proved to be a reliable method for making an intact sample assembly (Figure 4, 5) with tethers pre-tensioned and fully frozen within the ice, to allow for shear stress and simulation of pre-existing fault motions (Singh et al. 2019, 2021).

By following this procedure, the grain size of the ice surrounding the tether was kept uniform and reproducible, confirmed by analysis of images made by a Leica light microscope housed within a cold room (T= ~260 K). Images were collected both at the cross-section where the tether was present, and at adjoining ice grains. The ice produced by this method had an average grain size of 2.08 mm ± 0.22 as determined by the linear intercept method with a correction factor of 1.5 (Gifkins 1970). Additionally, four "control" samples were created by following the same protocols and die as above, but not including the tether.

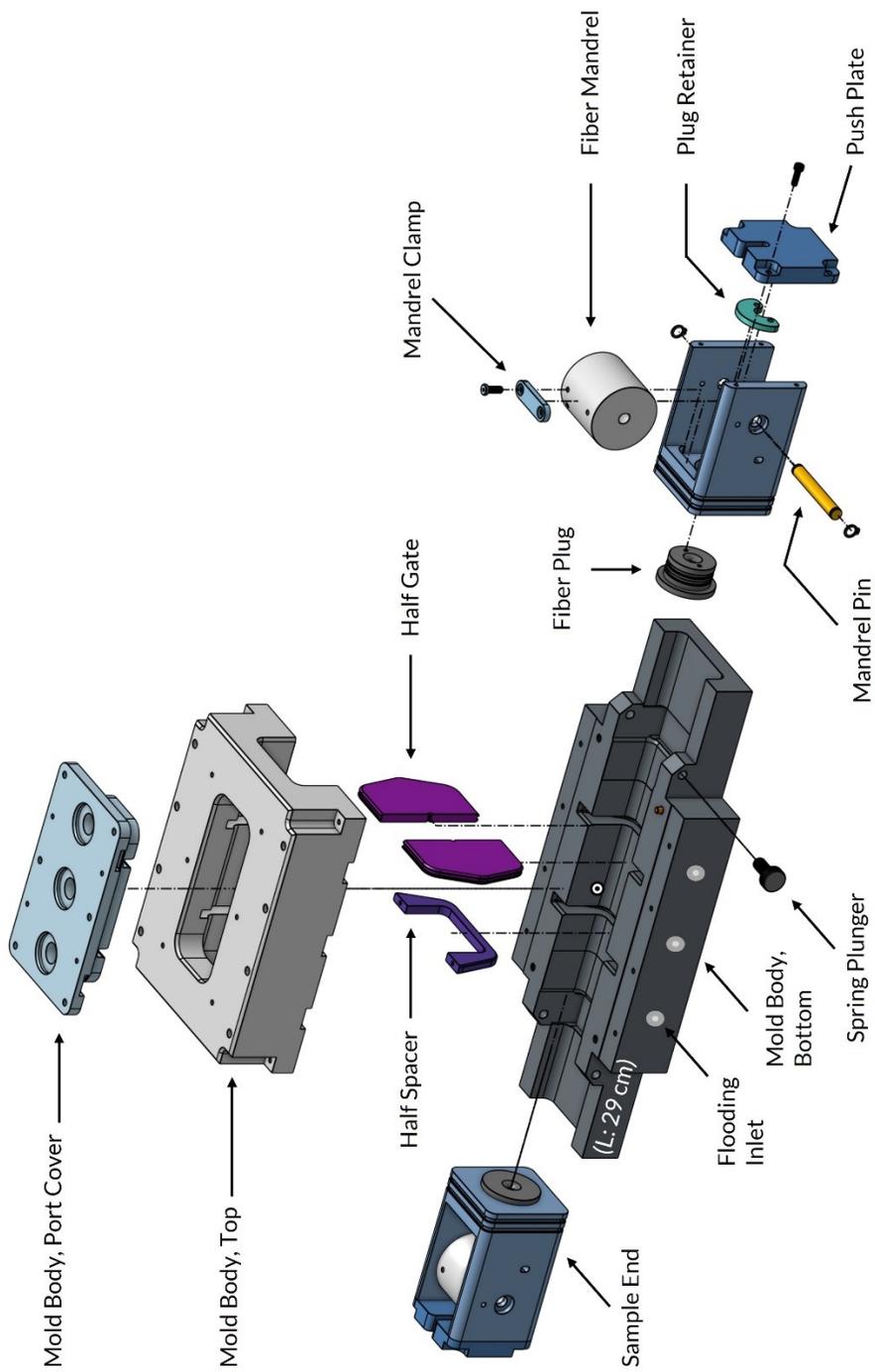

**Figure 3:** 3D model of Ice Fabrication Mold Components, custom built for freezing tethers in ice blocks under pretension.

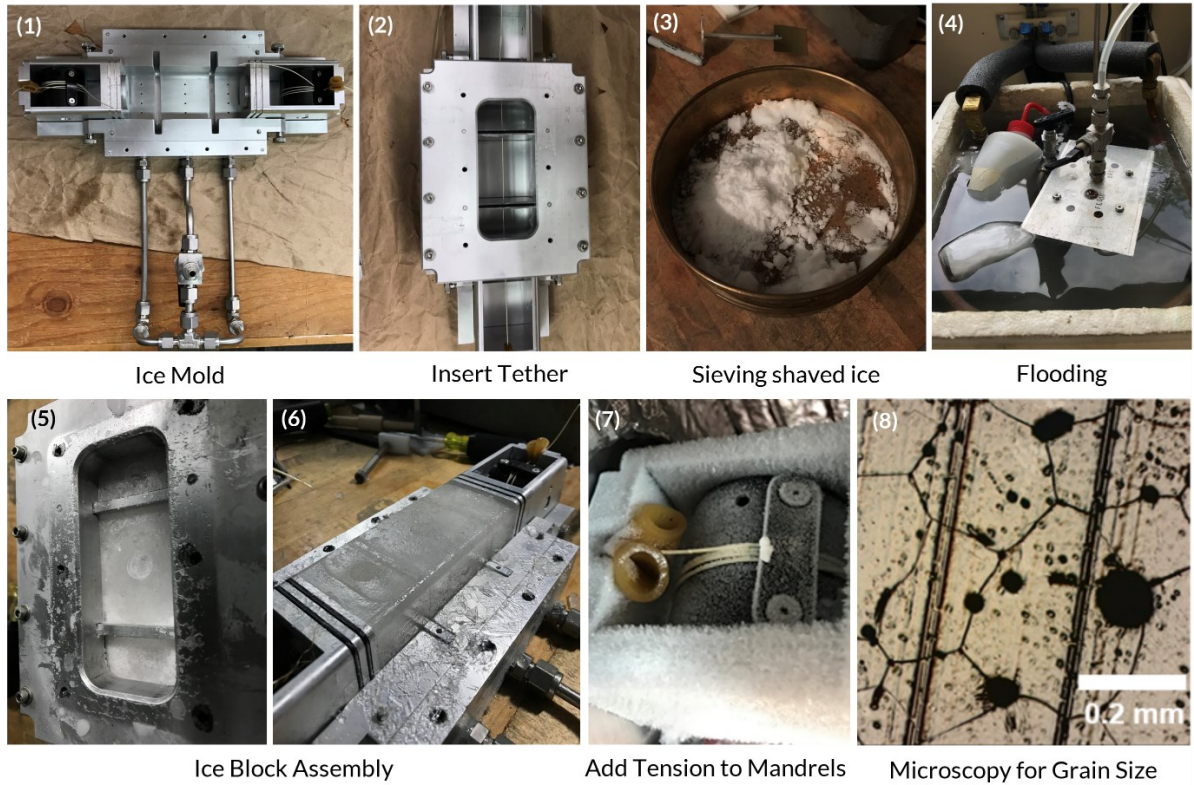

**Figure 4:** Ice sample fabrication protocol. Polycrystalline ice blocks of controlled grain size, porosity and impurity content are fabricated using a modified "standard ice" protocol (Cole, 1979) (1-5). Linden STFOC, HS STFOC tethers are embedded in ice (6-7), while images of ice away from tether (8) were used to characterize the grain size of the ice (uniform over the specimens tested).

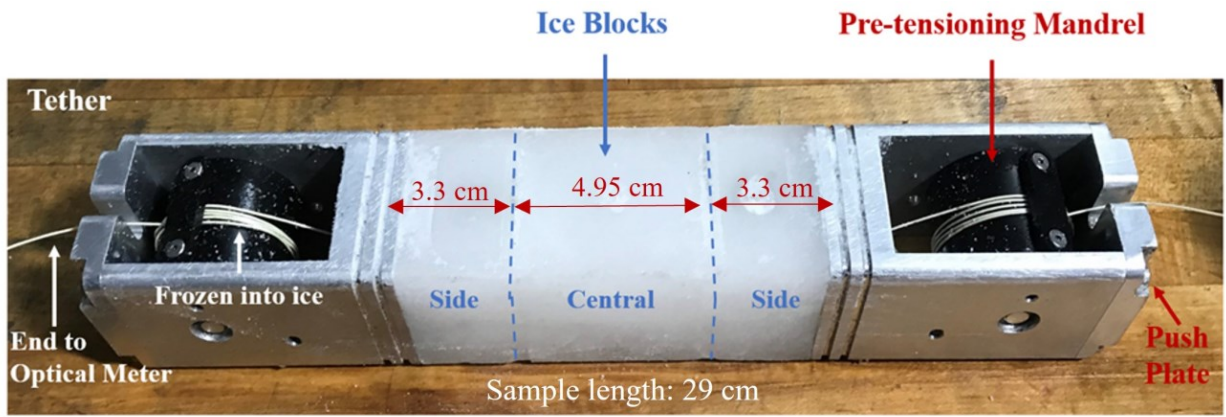

**Figure 5:** Custom sample assembly that allow**ed** pre-tensioned tethers to be frozen into ice. A small layer of lubrication between central (4.95×4.95×4.95 cm) and side blocks (3.30×4.95×4.95 cm), along with the geometry of forcing, causes the ice to break along the dash lines and shear the tether during loading.

### *3.3 Instrumentation and Test Conditions*
#### 3.3.1 Servo-Hydraulic Biaxial Deformation Apparatus

Shear tests were conducted using a cryogenic, servo-controlled biaxial deformation apparatus located at Lamont-Doherty Earth Observatory (LDEO), described in McCarthy et al. (2016), and shown in Figure 6 (top). We employed a modified double-direct-shear configuration, with the tether embedded through the middle of all three ice blocks. The side ice block samples sit on fiberglass support blocks bolted directly to the cryostat base to prevent rotational motion induced by direct shear, while maintaining an insulating layer (Figure 6 bottom). Molybdenum powder coating is used on the sides of the vertical ice/plastic forcer block to ensure frictionless sliding surface contact with the cryostat. Servo-controlled hydraulic pistons applied a normal stress in the horizontal direction while a velocity program controlled the downward vertical velocity, pushing the central sliding block past the outer stationary blocks. Three commercial load cells (Transducer Techniques, 454 kg maximum) and two direct current differential transformers (DCDTs; Schaevitz 2000 DC) located outside the cryostat measured vertical load (resolves as shear stress), horizontal load (resolves as normal stress), and displacement over time. Voltage signals for all measurements were collected by a 32-channel, 16-bit data acquisition system (National Instruments) at a sampling rate of 1 kHz per channel. The resolutions of load and displacement are 39 Pa and 0.7 micron, respectively. Shear stress is calculated over the two 50 mm x 50 mm sliding interfaces of the side blocks, while friction is calculated from the ratio of the shear stress to the normal stress.

Our experimental temperature range (100 – 260 K) was chosen to represent the full range of thermal conditions in the icy shell, excluding those conditions already tested in terrestrial applications. Temperature was maintained by a Liquid Nitrogen (LN)-cooled circulating chiller that pumped low temperature fluid (ethanol-water mixture) through copper channels next to the sample in a cryostat (Figure 6 bottom). For the lowest temperatures, the cryostat was directly cooled by LN flowing through the channels. Vacuum was used for thermal insulation, and temperature was monitored to ±0.1 K by Resistive Temperature Detectors (RTDs; Omega P-ULTRA) embedded in the walls of the cryostat adjacent to the sample. Low thermal conductivity ceramic pistons (Macor™) were used to translate stress from the hydraulic pistons to the sample ends, passing through Teflon seals in steel plates on the side panels of the cryostat. Data was logged for a selection of displacement speeds and temperatures across the ranges stated.

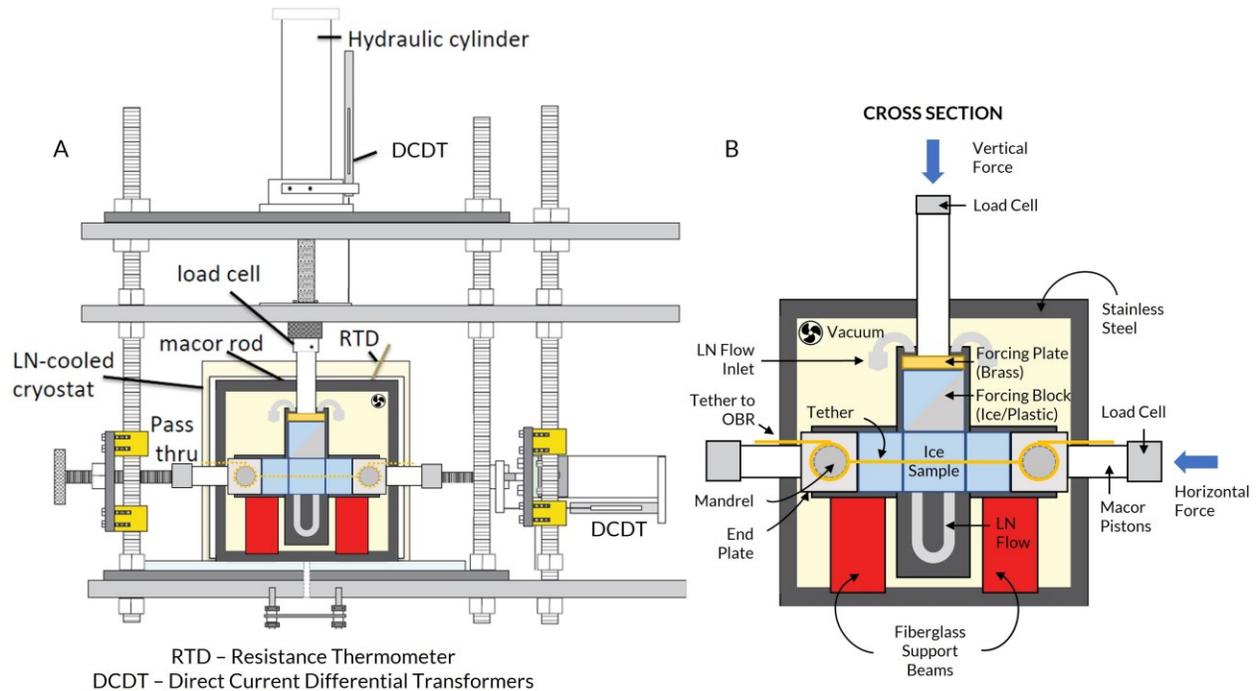

**Figure 6:** (A) LN-cooled biaxial apparatus at LDEO. Tethers were frozen into ice and loads applied as indicated by arrows. Ends of tethers were wrapped around mandrels and passed through the cryostat for monitoring for optical performance. (B) Schematic of the cryostat and load trains (vertical and horizontal) in cross section, showing the double direct shear configuration of the testing central ice sample and two stationary ice samples with/without tethers.

### 3.3.2 Shear Velocities and Driving Program

The ability of the tethers to withstand shear force is likely dependent on the rate of shearing. Here, we considered end member scenarios, as there are no known fault sliding rates for Europa. One scenario would be slow rates associated with creeping of a fault. Creeping strike-slip faults on Earth, which are driven by plate tectonics, slide at rates on order of $10^{-9}$ m/s. Europan faults, driven instead by tidal forcing, probably creep at higher rates, on order of $10^{-7}$ m/s (Nimmo & Gaidos 2002). On the other end of the spectrum would be a fault that is otherwise stuck but then lurches forward during a seismic event. The calculation of potential slip rates used here is discussed in Appendix A, and is on order of $10^{-7}$ to $10^{-4}$ m/s. For these tests, the sample assembly was pushed together under feedback-controlled normal stress (~100 kPa) and the central sliding block was forced down under controlled velocities that represent a potential range of fault sliding behavior on Europa and Enceladus (Nimmo et al. 2007; Sleep 2019; Rhoden et al. 2012), from creeping (ramps of 0.5, 1, and 10 μm/s) to seismic, stick-slip events (100, 200, and 300 μm/s; complete range of $3.0\times10^{-4}$ to $5.0\times10^{-7}$ m/s), each following a zero-velocity hold. Even without knowing Europa's slip rates exactly, these tests provide a measure of the robustness of tethers to shear.

Through a MATLAB routine, we create a simple LabView driving program that provides a voltage to the vertical servo-valve to drive the vertical piston cylinder forward at a designated speed for a designated time. The driving program consists of a series of velocity steps and holds (see Table 1 for details) to simulate the expected behavior of a pre-existing ice fault across which the path (and tether) of the cryobot passes, during the stages of (1) reactivation (loading); (2) long periods of creeping motion along fault planes followed by periods of inactivity (creeping to holds); and (3) sudden slips or ice quake events.

The displacement length for a single run is limited by the stroke of the vertical DCDT's linear range. Hence the total vertical displacement during testing was either 1.5 cm or 2.4 cm at each ice fracture, depending on the total number of driving velocity programs that were used. The strain range is calculated as 0.13 – 0.42 m/m, depending on the normalization to the cable length in the deformation zone: whether the displacements at the two fractures are accommodated over the entire tether length embedded in ice (11.55 cm), or half the length of tether (5.78 cm). For the 1.5 cm displacement case, we ran one program (ICE #1, an example of which is provided in Figure 7A), which included a combination of low velocity, creeping fault events at 0.5 to 10 microns per second, zero-velocity holds in between, and two high velocity shear events at 100 and 200 microns per second. For 2.4 cm displacement, at the end of the first program, the DCDT was re-zeroed to its lowest position and a new program was run consisting only of high velocity shear events (or ice-quakes) with zero-velocity holds in between; this combined program is referred to as ICE #2 (Figure 7B). The period of re-zeroing, and associated changes in recorded stress and displacement are eliminated during data processing. Examples of driving programs are provided in Figure 7.

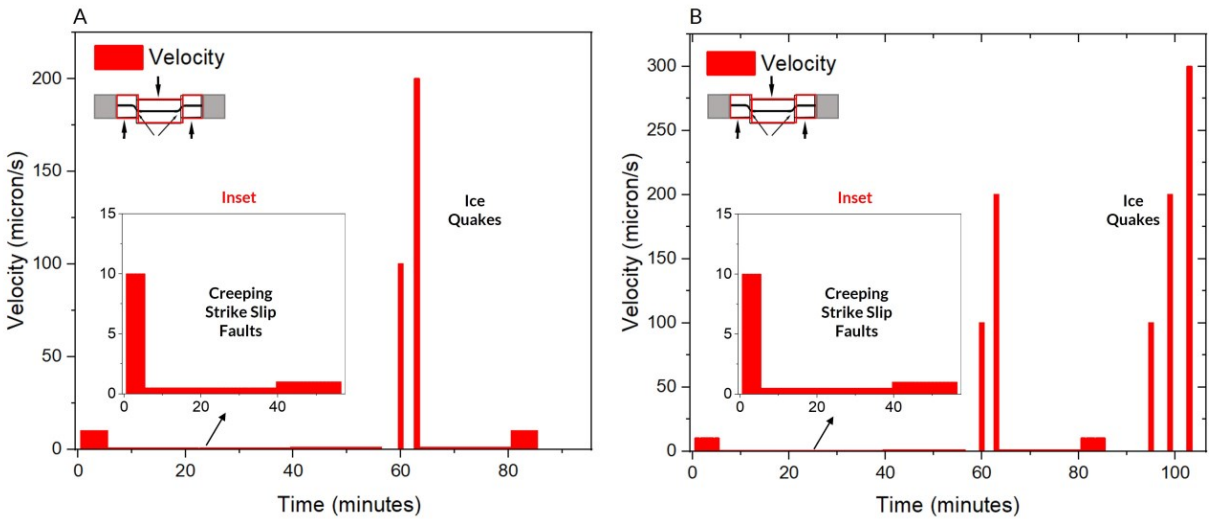

**Figure 7:** ICE#1 (A) and ICE#2 (B) velocity driving programs used during double direct shear tests. Velocities, durations, and displacements for each driving program are provided in Table 1.

**Table 1:** The velocity-control driving programs (ICE#1 and ICE#2) used for shear testing.

| Velocity (µm/s) | Time (s) | Shear Event | Displacement (mm) |
|---|---|---|---|
| 0 | 60 | Hold | 0.0 |
| 10 | 300 | Loading / Creeping | 3.0 |
| 0.5 | 2000 | Creeping Strike-Slip | 4.0 |
| 1 | 1000 | Creeping Strike-Slip | 5.0 |
| 0 | 200 | Hold (/ Relaxation) | 5.0 |
| 100 | 30 | Ice-Quake (Sudden Slip) | 8.0 |
| 0 | 200 | Hold | 8.0 |
| 200 | 15 | Ice-Quake (Sudden Slip) | 11.0 |
| 1 | 1000 | Creeping Strike-Slip | 12.0 |
| 10 | 300 | Creeping Strike-Slip | 15.0 |
| 0 | 200 | (Long) Hold | 15.0 |
| Additional ramps of Ice-Quakes added to ICE#1 program: full set referred as ICE#2 = ICE#1 + 3 quakes listed below ||||
| 100 | 30 | Ice-Quake | 18.0 |
| 0 | 200 | Hold | 18.0 |
| 200 | 15 | Ice-Quake | 21.0 |
| 0 | 200 | Hold | 21.0 |
| 300 | 10 | Ice-Quake | 24.0 |
| 0 | 200 | Hold | 24.0 |

### 3.3.3 Optical Source and Meter

The communications performance and viability of each tether was continuously monitored while inside the test apparatus. Pass-through ports on the side of the cryostat allowed the two ends of the tether to pass through to instrumentation. Optical insertion loss (IL) was continuously measured during testing using a stabilized light source (OptoTest OP250) at 1310 nm and optical power meter (OptoTest OP735), each instrument at one end of the cable. The relative change in signal (RIL) during testing was also calculated, based on ambient measurements of cables. To connect instrumentation, the tether was stripped to the cladding surrounding the fiber, cleaned and fusion spliced onto pre-terminated single-ended patch cables with connectors matching the instruments (FC/APC). We also monitored insertion loss prior to and during freezing of the tethers inside the solid blocks of ice, as controls, and compared to performance at the end of testing to characterize any degradation.

### 3.3.4 Microscopic Characterization

Characterization of representative ice samples after testing were conducted using a Leica DM2700 light microscope housed within a cold room (260 K), as described earlier in Section 3.2 for verifying homogenous ice-grain distributions. Prior to imaging, a mirror finish was obtained on the imaged surface using a microtome. Images were taken in refracted light and in most cases since the grain size was so large, were taken in only 2.5x magnification. Sublimation after microtoming accentuates grain boundaries. From these images, we used the linear intercept method to determine grain size. Additionally, a JEOL JCM-7000 NeoScope™ Benchtop Scanning Electron Microscope (SEM) was used to image the STFOC and HS STFOC tether samples post shear testing at 175 K and 150 K, respectively, using topographic (secondary electrons) and compositional mapping (backscatter electrons).

### 4. Results of Shear Testing

A total of 15 double-direct shear experiments were successfully conducted for pure water ice samples with embedded Linden STFOC and HS STFOC tethers using a cryogenic biaxial friction apparatus (Table 2). Mechanical properties of the tether-ice assemblage, such as peak shear stress achieved to activate faulting, shear stress vs velocity relationship, and optical insertion loss of tethers were measured. The experimental temperature range of 95 – 260 K is comparable to Europa's ice shell temperatures (from the near-surface to melting near the ice-ocean interface), with the velocity range of $5\times10^{-7}$ to $3\times10^{-4}$ m/s representing creeping faults and ice-quakes on Europa. Both tethers were tested at similar temperatures (e.g., 120 K and 150 K for HS STFOC versus 125-143 K for STFOC) to allow overlap in the expected mechanical response of the ice, while collectively covering the ice shell temperature range. Four control samples of pure water ice without tethers were also tested at ~100 K, 183 K, 210 K and 230 K for comparison close to the 2 extreme ice shell temperatures, and in the interior of the shell.

**Table 2:** Peak Stress for various samples and Driving Programs tested in order of increasing temperature/depth into ice shell.

| Temperature (K) | Tether Type | Driving Program | Peak Stress (kPa) | Ductile vs Brittle | File ID |
|---|---|---|---|---|---|
| 95-100 | Control | ICE#2 | 841 | Brittle | C00161 |
| 95-100 | STFOC | ICE#2 | 1288* | Brittle | C00148 |
| 120 | HS STFOC | ICE#2 | 1467 | Brittle | C00149 |
| 125-143^ | STFOC | ICE#2 | 951 | Brittle | C00291 |
| 150 | HS STFOC | ICE#2 | 893 | Brittle | C00259 |
| 175 | STFOC | ICE#2 | 830 | Brittle | C00281 |
| 183 | Control | ICE#2 | 655 | Brittle | C00255 |
| 195 | STFOC | ICE#1 | 1061* | Brittle | C00142 |
| 199 | HS STFOC | ICE#1 | 1061* | Brittle | C00139 |
| 198 – 233^ | STFOC | ICE#1 | 1061* | Brittle | C00137 |
| 208-213 | Control | ICE#2 | 1924* | Brittle | C00257 |
| 220 | HS STFOC | ICE#1 | 1201 | Brittle | C00147 |
| 230 | STFOC | ICE#1 | 1061* | Brittle | C00133 |
| 230 | Control | ICE#2 | 1548 | Brittle | C00158 |
| 237 | HS STFOC | ICE#2 | 550 | Ductile | C00150 |
| 248 | HS STFOC | ICE#1 | 179 | Ductile | C0144 |
| 255 | STFOC | ICE#1 | 755 | Ductile | C0291 |
| 260 | HS STFOC | ICE#1 | 296 | Ductile | C0061 |
| 260 | STFOC | ICE#1 | 67 | Ductile | C0060 |

^Peak stress was achieved at the initial set point (143 K and 233 K respectively)
*Data was initially clipped at 1061 kPa due to calibration limits of vertical load cell. Load cells were later recalibrated.

*4.1 Initial Peak Stress for Ice Fault Activation: Implications for Ice Strength*

The initial strength of the ice block-tether assemblage until cracking at interfaces depends primarily on the strength of ice and as such depends on temperature. Hence, while it is valuable to report the peak stresses survived by the samples at the start of shear testing, those stresses provide more of an understanding of the ice behavior at a range of ice shell temperatures, rather than the limitations of tethers.

The ice-ice interface activates at a peak stress of ~70–300 kPa at 260 K (Figure 8). The STFOC-ice sample 'faults' are activated at the lower peak stress of 67 kPa, while the HS STFOC-ice sample requires a higher peak stress of 296 kPa for activation. This trend is consistent throughout testing down to 95 K, with the HS STFOC-ice samples offering more resistance to activation. A loud audible pop is heard at the instant of the ice-ice interface activation, with a significant drop in shear stress and the interfaces becoming visible through the cryostat windows. With decreasing temperature, the initial peak stress required increases to ~550 kPa (HS STFOC) at 237 K to ~1061* kPa (HS STFOC) at 230 K, which is the temperature at which we observe a phenomenological transition from ductile to brittle behavior. At temperatures colder still, the peak stress required jumps to 1201 kPa (HS STFOC) at 220 K. For a few experimental runs (HS STFOC at 199 K and STFOC at 195 K, 200 – 230 K; Figure 9B), the data were clipped at 1061 kPa due to the resolution of the vertical load cell that was calibrated for greater accuracy at lower stresses. We denote the clipped data with an asterisk in Figure 8. The vertical load cell was subsequently recalibrated to broaden the recording range for these high stress tests. For the control samples (ice only, no tethers), the recorded peak stress was ~1.55 MPa at 230 K, which is ~0.3 to 0.5* MPa (*data was clipped) higher than the recorded peak stress for the ice-tether samples at similar temperatures, and are comparable at the coldest temperatures tested (~120 K).

At intermediate temperatures (125 K to 175 K), the peak stress decreases to the range of ~830 to 951 kPa for the ice-tether samples, while the control sample (183 K) had a lower peak stress of ~655 kPa. At the lowest temperatures of testing (95 K to 120 K), the peak stresses increased further, up to ~ 1.3 MPa (STFOC) at 100 K to ~1.5 MPa (HS STFOC) at 120 K, demonstrating a roughly linear temperature dependence of ~10.5 to 11 kPa/K (linear fits of HS STFOC and STFOC ice-tether samples). The control sample at 100 K, again has a lower peak strength of ~841 kPa, which is comparable to tether-ice sample strength at intermediate temperatures. The only difference between the two sample types is the lack of a tether in the control sample indicating that the tether is potentially offering resistance to shear. Overall, the peak stresses documented here (Figure 8, Table 2) represent the activation of the interfaces, with the tether-ice samples experiencing peak stresses of the order of ~1.3 to 1.5 MPa at the coldest temperatures tested.

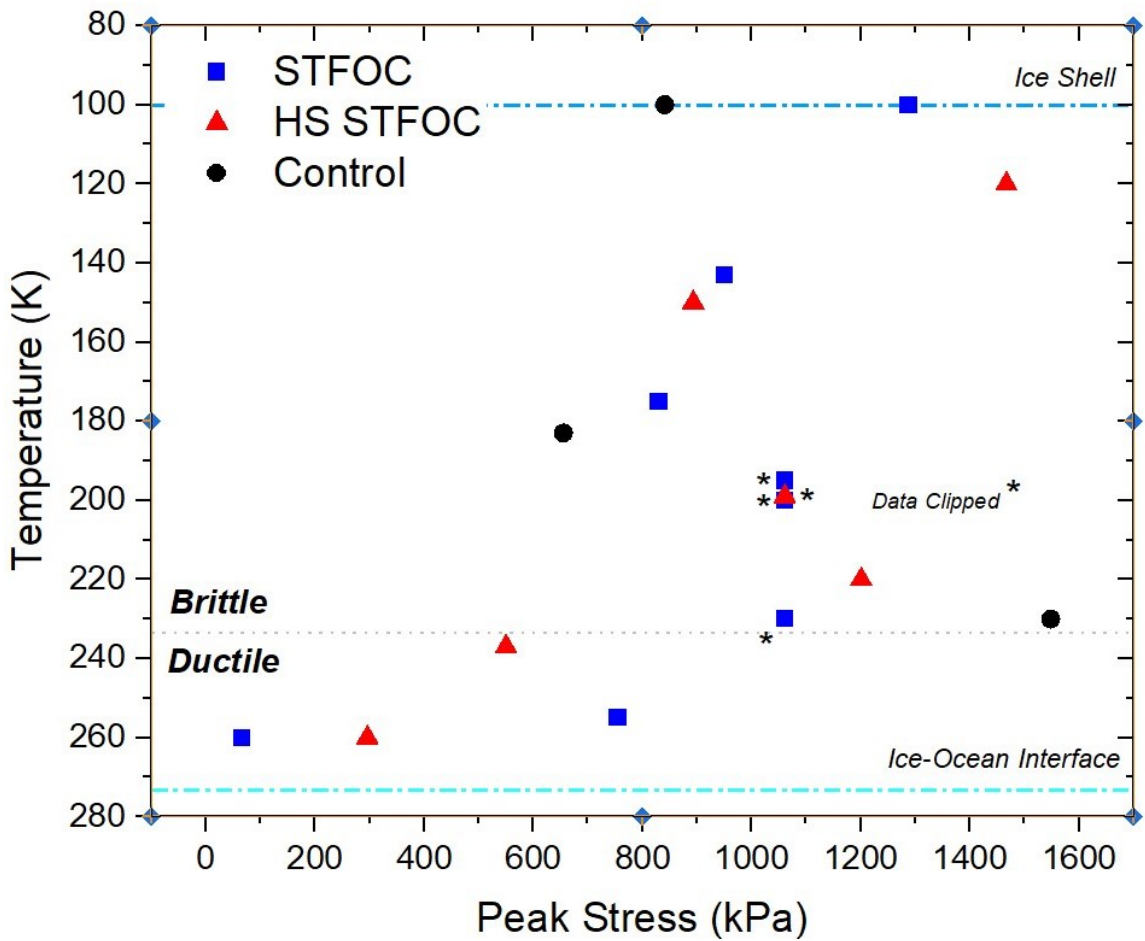

**Figure 8:** Peak stress survived by STFOC and HS STFOC tethers in ice and control ice sample over the range of ice shell temperatures. *where indicated data was clipped at 1061 kPa due to calibration limits of vertical load cell, thus representing an underreporting. Load cells were later recalibrated to fix this issue.

*4.2 Shear Stress: Velocity versus Temperature Relationship*

Figure 9 shows shear stress (black) and velocity (red) versus time for the STFOC tether tested at 230 K demonstrating a representative response to the applied velocity driving program(s). Appendix B discusses a classical example of frictional behavior of ice, including the dependence of shear stress on a change in sliding velocity as demonstrated here, along with definitions of the terms discussed in this section. The initial increase in shear stress from zero occurs prior to the sample ice-ice interface activation and represents the stiffness of the apparatus and sample assembly. In a typical friction test, this value increases only to the steady-state friction value and then levels off. However, in these experiments the ice-ice interfaces have been welded intact by the two-step freezing protocol. The peak stress after the initial ramp is due to the interfaces activating/breaking, as described in Section 4.1. In this test, the two interfaces on either side of the central block broke at separate times, thus the double peak (the first of which has clipped data), in Figure 9B, with a very large stress drop in between. In other tests, we observe a single peak if the two interfaces activate at the same time. The spike in the red velocity data in Figure 9A at ~75 minutes was not part of the programmed routine; it represents the piston lurching forward in response to the first interface activating. The downstep in velocity (as described in the program in Table 1) at about 77 minutes is superimposed on the relaxation and not evident in this dataset, but the small step up in velocity (from 0.5 to 1 μm/s) is seen at 110 minutes. That the shear stress response is gradual and hill-like suggests that the sliding interface contains some ice gouge from the initial interface activation (e.g., Chester 1994). The relatively smooth data at these two velocities are clear examples of stable sliding at these conditions.

      At approximately 128 minutes in Figure 9A, the piston was stopped, and classic relaxation of the apparatus and interfaces was observed. Following this, a 100 μm/s event was imposed by the routine. The response of the system was a series of stress drops (~300 kPa), with the piston lurching forward at each drop, as shown in the red velocity data (Figure 9C). That the stress returned to similar values after each drop indicates that the interfaces were still somewhat intact. This contrasts with the response to the 200 μm/s event at ~135 minutes. Here the stress drops were smaller (~100 kPa) spaced closer in time, and did not return to previous values, causing an overall decline in the shear stress. This indicates decoupling and damage at the ice interfaces. The gradual increase of shear stress in response to the velocity step at ~138 minutes further implies this. During both high velocity periods, the jagged response in the friction data is representative of unstable, stick-slip behavior in the system at these conditions. Audible pops were heard at each stress drop. Similar behavior was observed in the tether-less control ice sample at a similar temperature, so the stress drops are due to the ice and central plastic forcing block sliding and/or ice cracking, and not due to the tether.

      All other tests demonstrate some variation on the above description due to the temperature and sliding behavior of the ice, and the inferred character of the interface (i.e., intact versus gouged). In the following section, we will describe the variations in sliding with temperature and velocity, demonstrating the striking range of conditions that the tether experienced during testing (Figure 10 and 11).

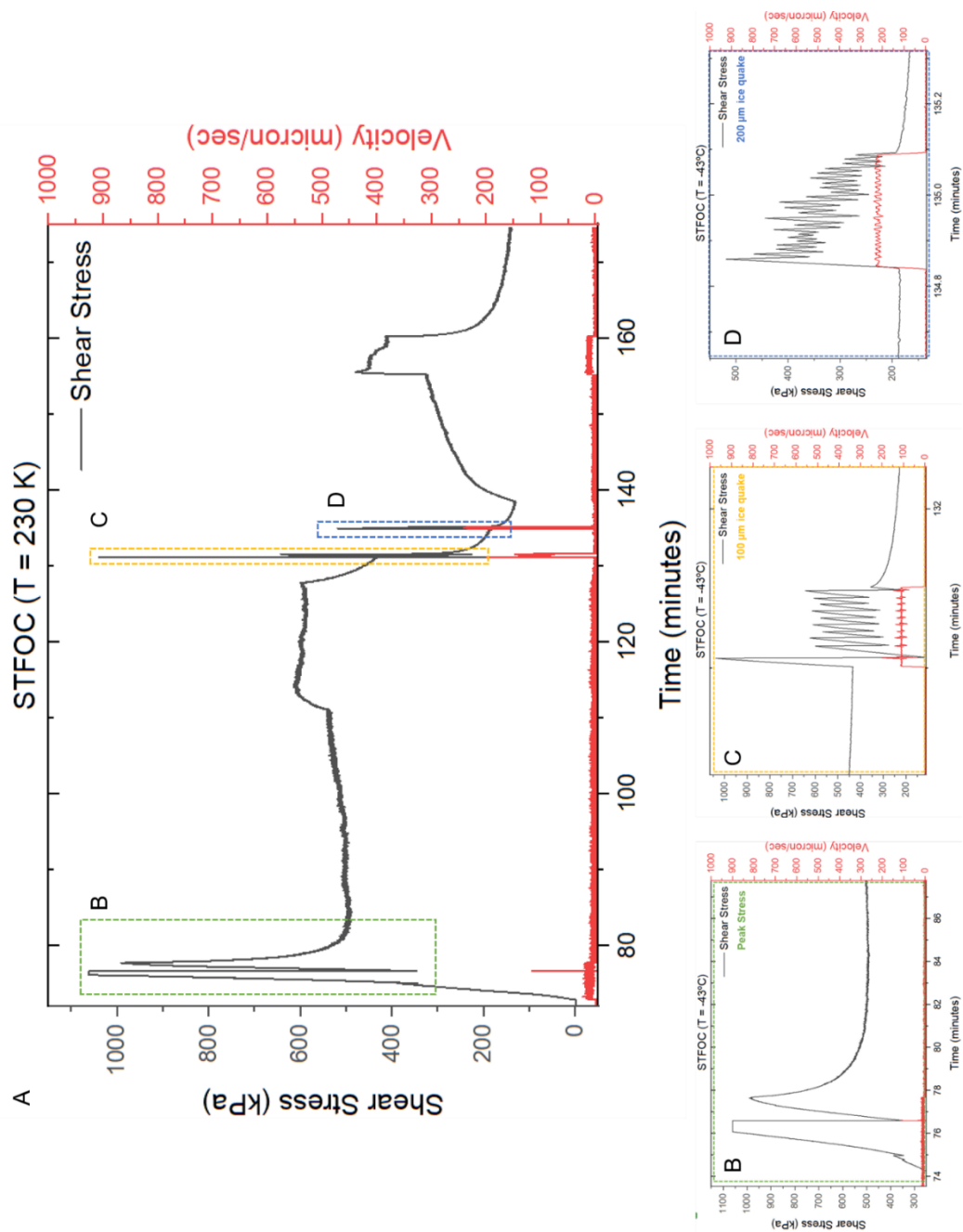

**Figure 9:** (A) Representative shear stress and velocity versus time curve for the STFOC tether at 230 K, highlighting (B) peak stress achieved; (C), (D) are ice quake events.

*4.3 Shear Stress versus Displacement Curves: Implications for Europa*

Figure 10 and 11 show shear stress curves versus displacement for each of the temperature-velocity combinations tested for each of the tethers tested. The curves indicate that for both tether assemblies at low velocity ($5\times10^{-7}$ to $1\times10^{-5}$ m/s) and high temperature (200 to 250 K), the frictional response is generally smooth and steady. At 250 K, the warmest temperatures studied, stable sliding is observed for the entire velocity range tested. On cooling to around 220 – 230 K, we observe a transition (for both tethers tested in Figure 10 and 11) to unstable, stick-slip sliding behavior at the highest velocities ($1\times10^{-4}$ m/s to $3\times10^{-4}$ m/s).

At intermediate temperatures (175 – 199 K), large stress drops with multiple stick-slips observed at low to intermediate velocities ($5\times10^{-7}$ to $1\times10^{-5}$ m/s), with highest frequency of these events observed at $1\times10^{-5}$ m/s. At high velocities ($1\times10^{-4}$ m/s to $3\times10^{-4}$ m/s) the system transitions from unstable stick-slip behavior to stable sliding. Both the STFOC and HS STFOC samples experience multiple repeatable drops of 40-80 kPa in shear stress during the loading phase. During follow up velocity ramps of 0.5, 1 to 10 μm/sec, the shear stress stabilized around an average of ~900 kPa, with consistent stress drops 90 to 140 kPa.

At the coldest conditions (100 – 150 K), the system transitions back to stable sliding at higher velocities, from unstable, stick-slip behavior at slower velocities. These results are nearly identical to those seen in a pure ice-on-ice (tetherless) friction study at comparable conditions (Schulson & Fortt 2012, their Figure 3). To validate this behavior, we evaluated four control samples in a range of temperatures covered for the tethers, from the coldest (at 100 K) to intermediate (at 183 K) and relatively warm (at 230 K). These control ice experiments demonstrated identical sliding behavior with velocity and temperature to the tethered runs, as illustrated in Figure 12, with implications discussed in Section 5.2.

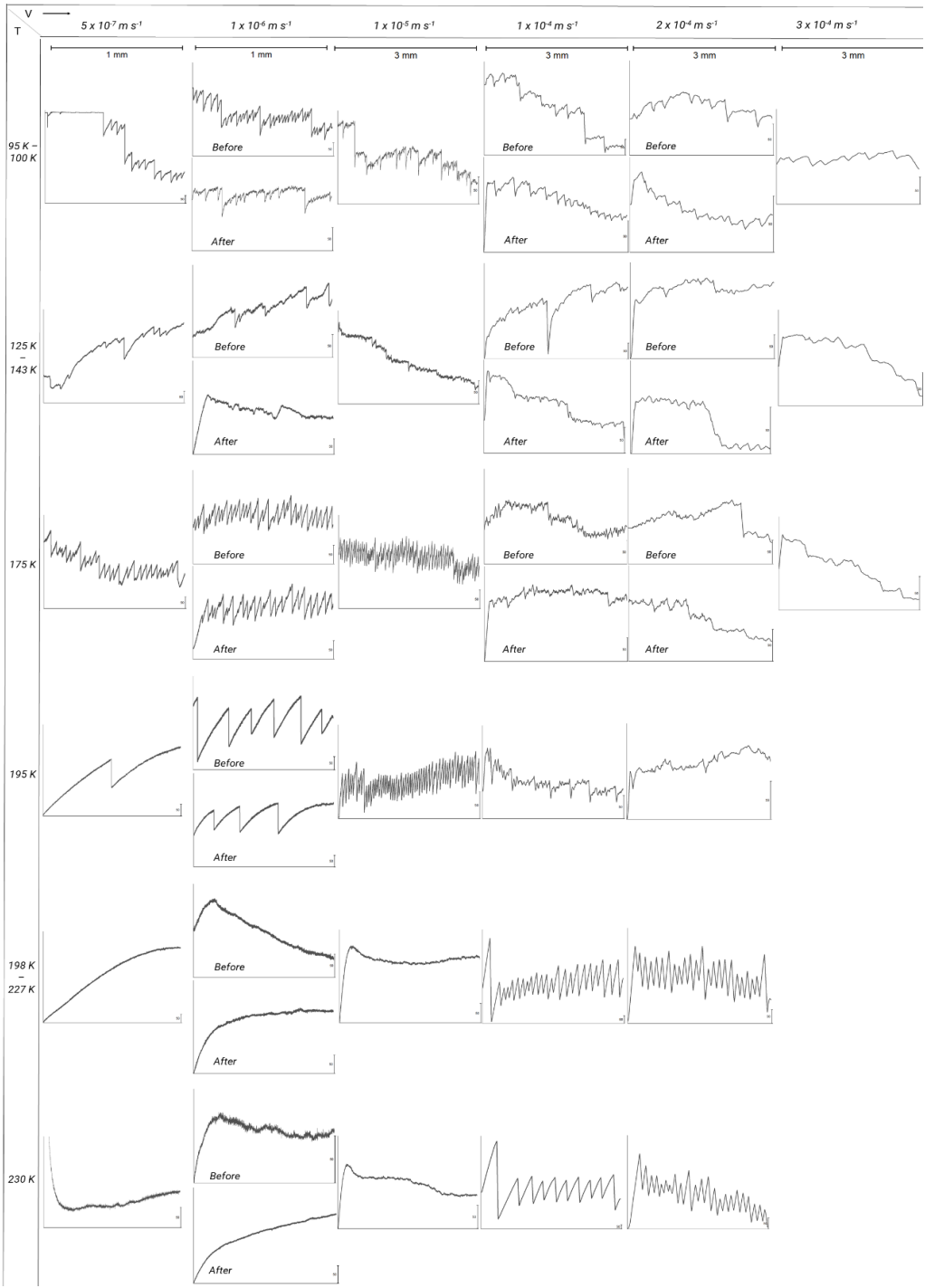

**Figure 10:** Shear stress curves (scale bar: 50 kPa) vs displacement (scale bar: 1 mm for runs at $5\times10^{-7}$ m s$^{-1}$ to $1\times10^{-5}$ m s$^{-1}$; 3 mm for $1\times10^{-4}$ m s$^{-1}$ and $2\times10^{-2}$ m s$^{-1}$) of STFOC tethers for temperature-velocity combinations tested under an applied normal stress of 100 kPa. The before and after labels indicate the condition of the central ice block: before is intact, while after is composed of gouge (due to cracking during high velocity events).

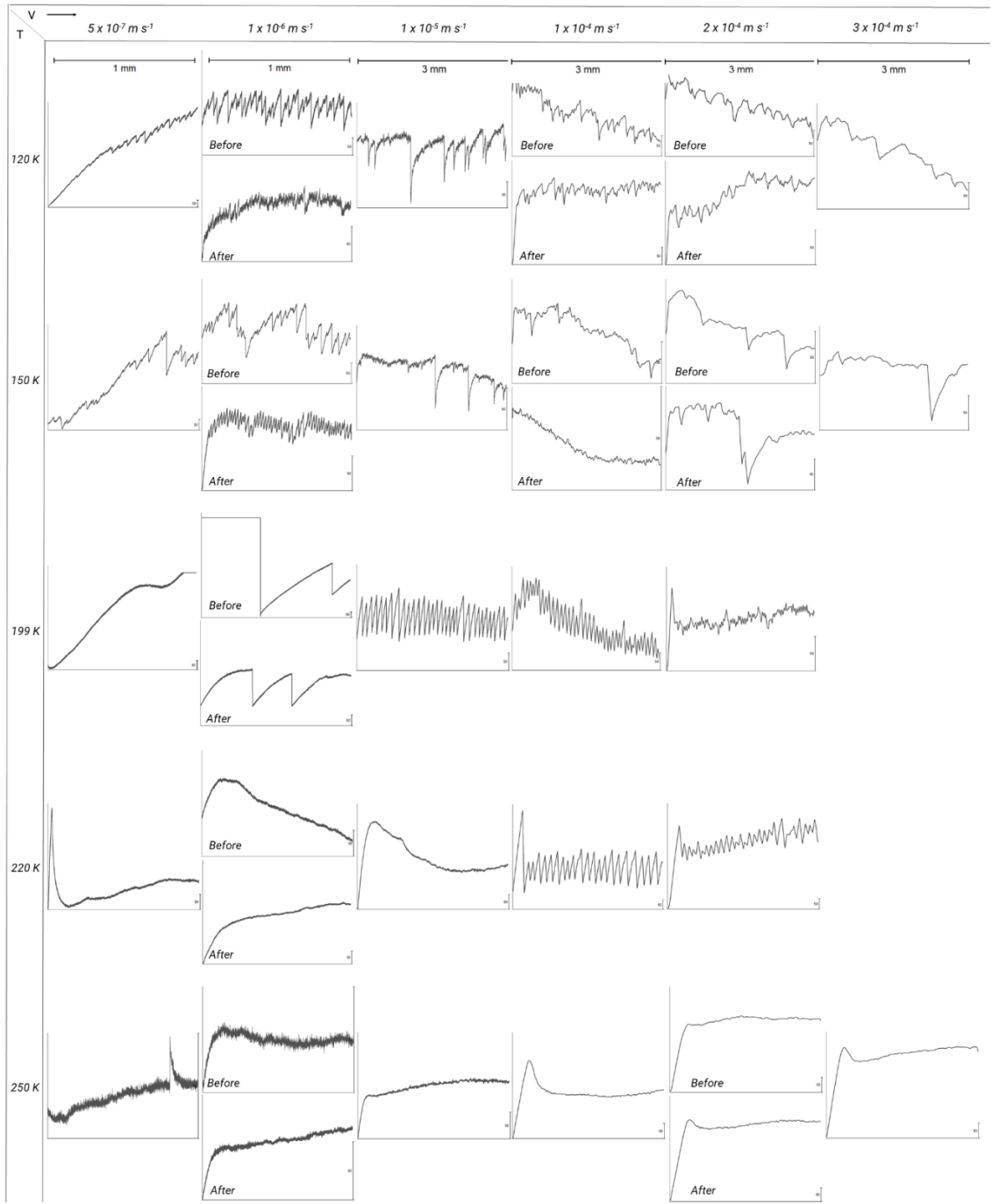

**Figure 11:** Shear stress curves (scale bar: 50 kPa) vs displacement (scale: 1 mm for runs at $5\times10^{-7}$ m s$^{-1}$ to $1\times10^{-5}$ m s$^{-1}$; 3 mm for $1\times10^{-4}$ m s$^{-1}$ and $2\times10^{-2}$ m s$^{-1}$) of HS STFOC tethers for temperature-velocity combinations tested under an applied normal stress of 100 kPa. The before and after labels indicate the condition of the central ice block: before is intact, while after is composed of gauge (due to cracking during high velocity events).

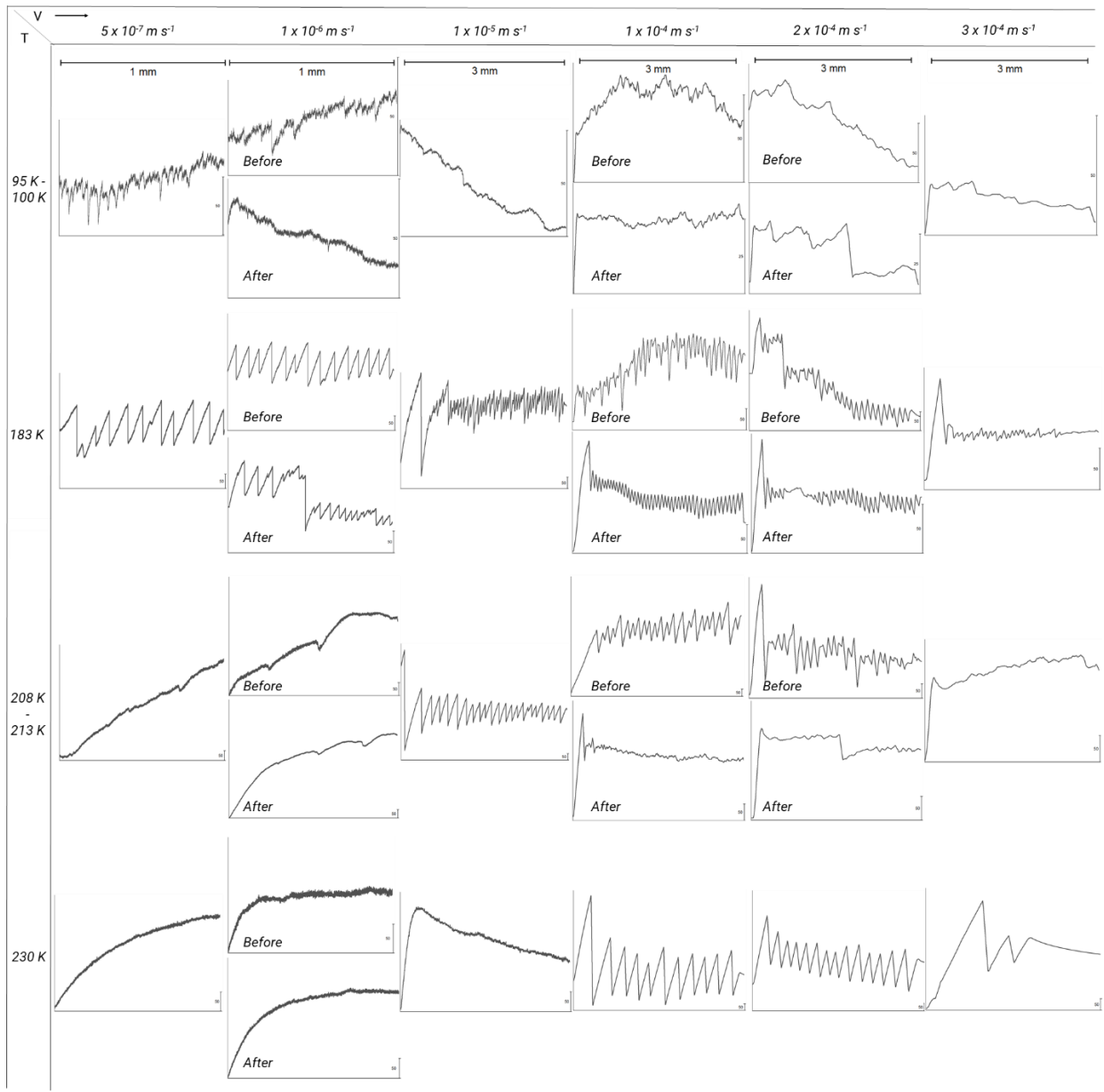

**Figure 12:** Shear stress curves (scale bar: 25 or 50 kPa) vs displacement (scale: 1 mm for runs at $5\times10^{-7}$ m s$^{-1}$ to $1\times10^{-5}$ m s$^{-1}$; 3 mm for $1\times10^{-4}$ m s$^{-1}$ and $2\times10^{-2}$ m s$^{-1}$) of control samples of ice for temperature-velocity combinations tested under an applied normal stress of 100 kPa.

*4.4 Behavior of Ice: Brittle to Ductile Transition Temperature and Damage to Central Ice Sample*

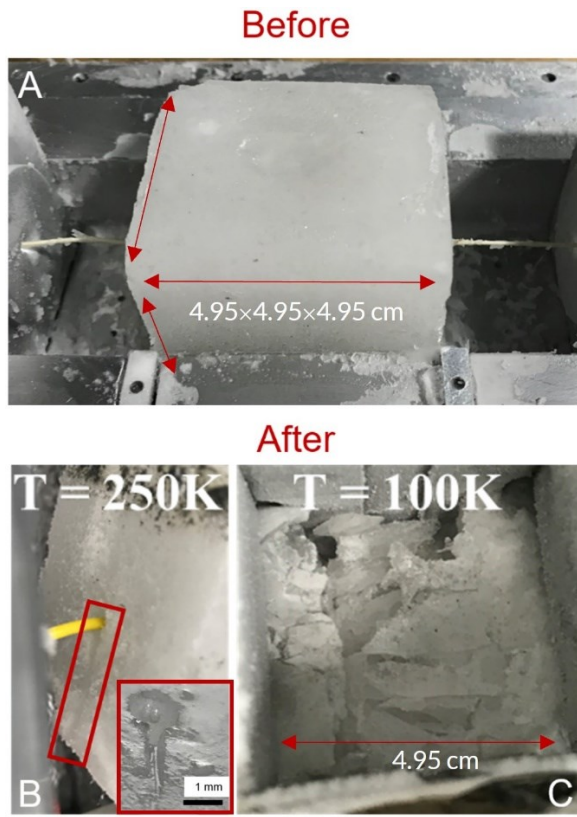

**Figure 13:** The behavior of the (A) central ice sample during shear testing depends on the temperature of the ice: (B) It behaves viscously at warmer conditions (left) and the tether cuts a groove (see inset); (C) at colder temperatures, the ice is pulverized.

The ice behavior relative to the embedded tether during shearing greatly depended on temperature. At warm temperatures, the ice behaved viscously allowing the tether to cut a groove through the ice, with the central ice block sliding down relative to the side blocks (Figure 13 B). In contrast, cracking and ultimately pulverizing of the ice was observed at the coldest temperatures, with cracking predominantly observed at the high shearing rates of 100, 200 and 300 μm/s (Figure 13 C,). This phenomenological brittle to ductile transition occurred at around 230 K.

*4.5 Macroscopic Damage to Tether*

Both the STFOC and HS STFOC tethers tested in this study survived shearing across pre-existing 'ice-faults' in a pure water ice shell (Figure 14). At intermediate to warm temperature conditions (195-260 K), the tethers remain largely undeformed, finishing in a u-shape after testing (Figure 14A). Only at the coldest temperatures (~ 100 K) and highest velocities tested ($3\times10^{-4}$ m/s), did we observe that the LCP strength jacket of the STFOC tether broke, revealing the intact optical fiber inside. The HS STFOC tether also survived at the coldest temperatures, with only a minor cut to the outer polyurethane jacket and a kink at the shear interface (Figure 14E). No damage was observed to the Kevlar braid, LCP strength member, or the optical fiber inside. At these temperatures, the ice was cold (and hard) enough to partially prevent any movement of the tether through the ice to accommodate the vertical stress. However, the tethers proved robust to these conditions, which increased confidence in the HS STFOC tethers for potential application at Ocean Worlds.

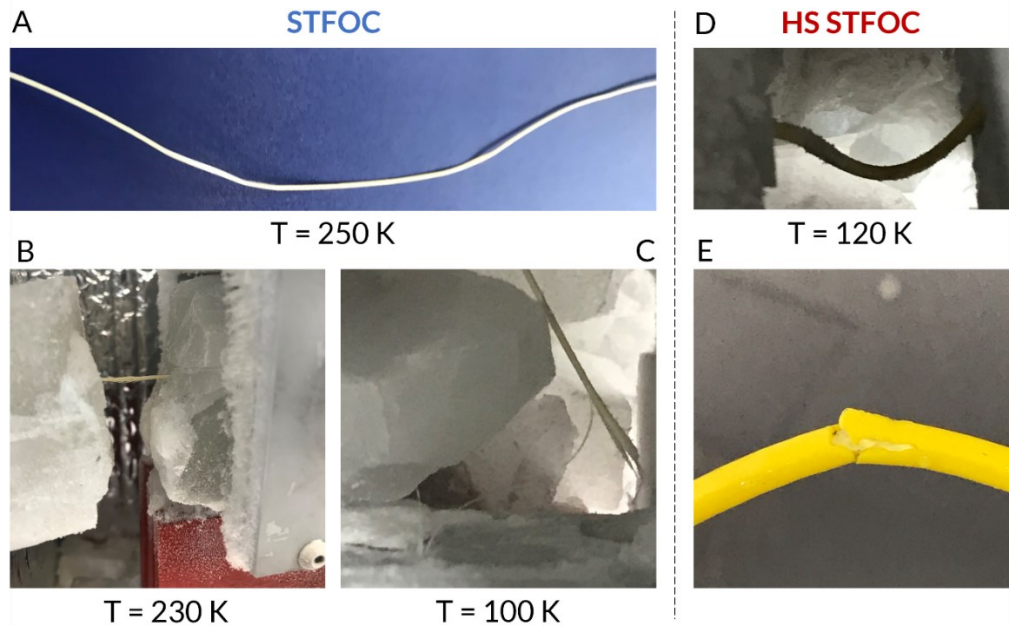

**Figure 14**: Damage to the outer layer of STFOC (A to C) and HS STFOC (D-E) during testing to accommodate shearing, with kinks observed at lower temperatures (T = 250 K) and low velocity runs (D).

*4.6 Microscopic Characterization of Ice Samples After Testing*

Exemplars of both warm (260 K) and cold (195 – 199 K) ice samples were evaluated to constrain the grain size of the ice (Figure 4(8)), as described in Section 3.2.1, and identify deformation patterns. Consistent with the macroscopic observations of ductile behavior of ice at warm temperatures (250 – 260 K) described in Section 3.1, microscopic grooves cut by tethers are visible at the interfaces and around the center of the sample housing the tether, with a surrounding flow-like pattern due to potential melting and refreezing of the ice (Figure 13B). At colder temperatures, with progressi**ve** cracking of the central blocks of ice and macroscopic observations of brittle behavior (from 230 K to 100 K), microscopic fractures were visible in both the central and side blocks (Figure 15A).

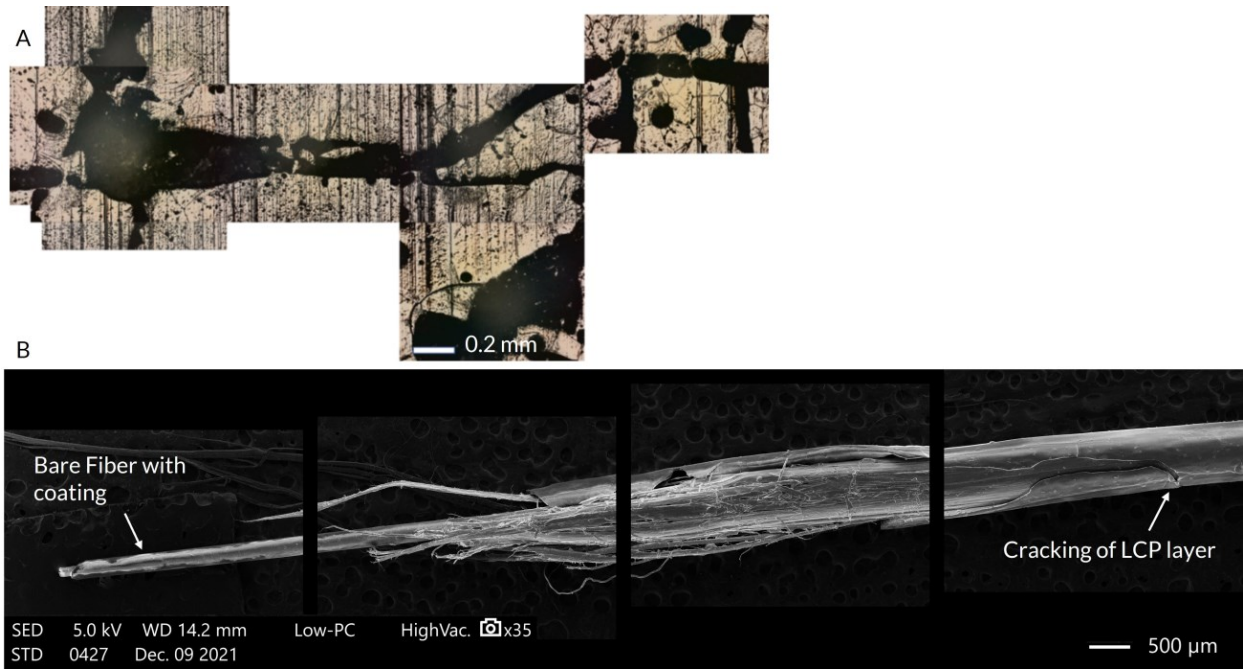

**Figure 15:** (A) Fracture network formed inside ice sample post shear testing at 199 K; STFOC fiber (175 K) imaged with a Scanning Electron Microscope (SEM), using secondary electrons (B) shows the failure of the outer LCP layer (bottom right panel) and impact on bare fiber (bottom left panel).

Microscopic characterization of post-tested samples revealed a large network of fractures pervading the ice samples (Figure 15A), which likely lowered the strength of the ice and allowed movement of the tether through the surrounding ice, resulting in survival at colder temperatures. Preliminary analysis with the Scanning Electron Microscope (SEM) on the STFOC tether is demonstrated in Figure 15-bottom, which highlights that the SEM can serve as an important tool for characterization of failure modes of tethers. Here, the larger diameter of the HS STFOC outer jacket (OD 1.9 mm) made it unsuitable for high resolution analysis due to lack of focus. Images of the STFOC fiber in Figure 15B (post testing at 175 K) display the cracking of the LCP layer and damage to bare fiber inside, close to the point of failure.

*4.7 Optical Signal Loss Characterization*

Optical signal and transmission loss for both Linden tethers were recorded, using the procedure mentioned in the Section 3.3.3 (Table 3). The optical meter reports higher values (e.g., 1-2 dB) to indicate some optical insertion loss during testing, and lower values (e.g., -70 to -98 dB) to indicate significant insertion loss such that no communication would be possible with standard hardware. Loss is determined by the relative change in signal (RIL) during testing.

**Table 3:** List of samples and optical measurements (OM). Yellow columns highlight STFOC sample tests with high initial optical insertion loss (IL); Orange column highlight no measurements.

| Temperature (K) | Tether Type | Peak Stress(kPa) | OM (Y/N) | IL (dB) | Max RIL (±dB) | File ID |
|---|---|---|---|---|---|---|
| 95-100 | STFOC | 1288* | Y | -89.1 | 0.5 | C0148 |
| 120 | HS STFOC | 1467 | Y | -32.3 | 6.6 | C0149 |
| 125-143^ | STFOC | 951 | Y | -89 | - | C0291 |
| 150 | HS STFOC | 893 | Y | -66.8 | 6.4 / -4.4 | C0259 |
| 175 | STFOC | 830 | Y | -98.1 | - | C0281 |
| 195 | STFOC | 1061* | Y | -82.3 | -6.8 | C0142 |
| 199 | HS STFOC | 1061* | Y | -68.7 | 5.9 / -7.0 | C0139 |
| 198 – 233^ | STFOC | 1061* | Y | -89.1 | - | C0137 |
| 220 | HS STFOC | 1201 | Y | -74.3 | 4.7 | C0147 |
| 230 | STFOC | 1061* | N | - | - | C0133 |
| 237 | HS STFOC | 550 | Y | -31.0 | -0.1 | C0150 |
| 248 | HS STFOC | 179 | Y | -65.1 | 0.5 / -5.9 | C0144 |
| 255 | STFOC | 1451 | Y | 0.47 | -2.57 | C0291 |
| 260 | HS STFOC | 296 | N | - | - | C0061 |
| 260 | STFOC | 67 | N | - | - | C0060 |

#### 4.7.1 STFOC Tether Signal Loss

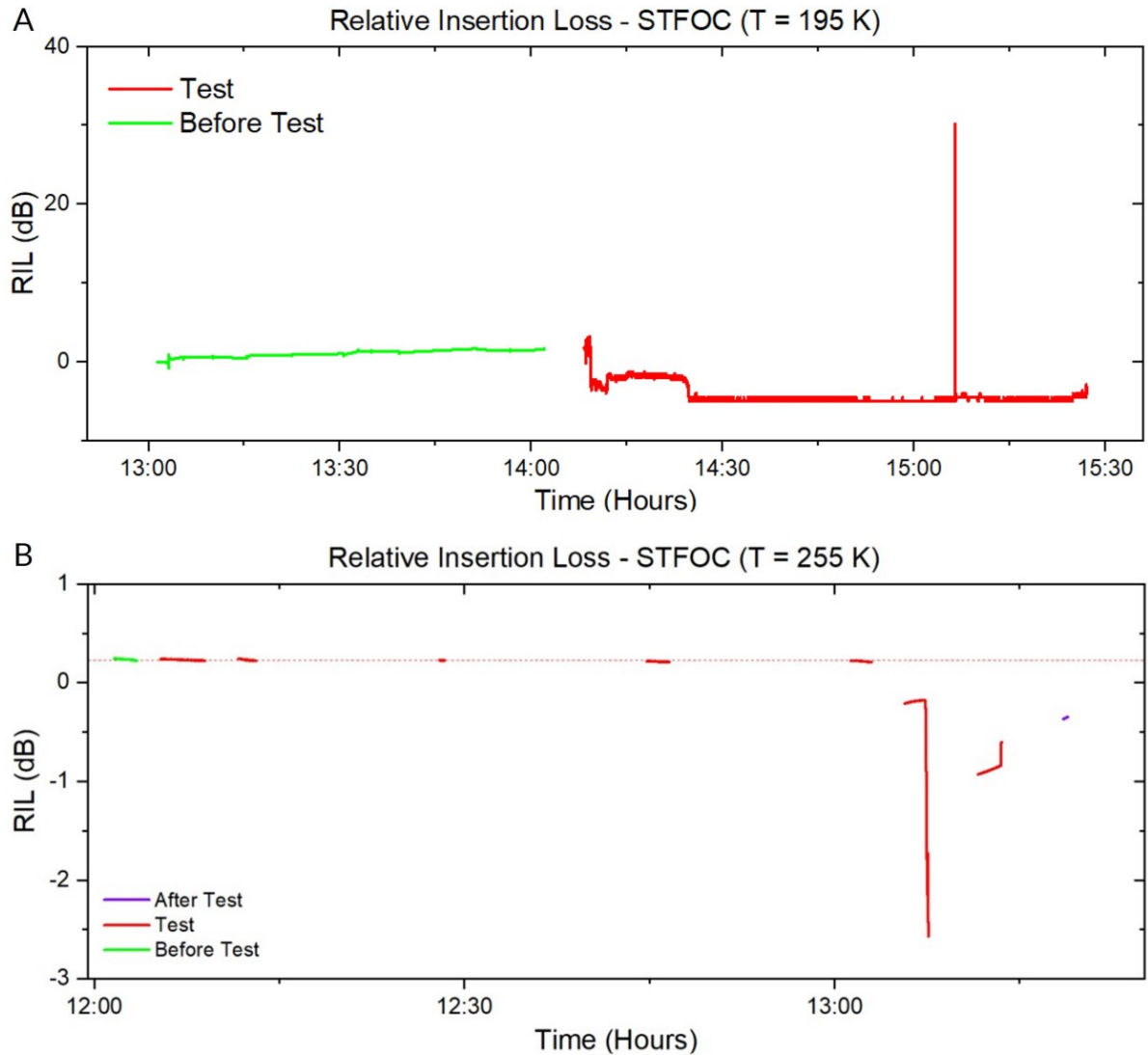

**Figure 16:** Relative Insertion Loss recorded for STFOC tether at (A) 195 K, and (B) 255 K.

We attempted to collect the optical signal for STFOC tethers at six temperature ranges of: (A) 95-100 K, (B) 125-143 K, (C) 175 K, (D) 195 K, (E) 198-233 K, and (F) 255 K. Unfortunately, the first 5 measurements show significant insertion losses of -82 dB to -98 dB post freezing of tether in ice samples and fusion splicing (i.e., prior to start of testing). The highest recorded optical signal (approximately -82 dB) at 195 K is still deemed insufficient to enable communication as the baseline signal lies near the noise floor (Figure 16). The source of this high insertion loss can either be due to (i) crimps/breaks in the fiber during loading inside the ice fabrication mold for pre-freezing in the cold room, or (ii) issues in fusion splicing of tether to connectors. The first hypothesis is supported by the observed macroscopic damage to tethers, discussed in Section 4.5, while the second hypothesis is negated to a degree by results from Section 4.7.2, where identical fusion splicing procedure was followed for HS STFOC tethers, and the optical signal was successfully recorded. As stated earlier, the signal is insufficient to evaluate, and subsequent velocity driving program runs at 195 K result in a relative insertion loss of ~ 7 dB (down to -90 dB) implying complete loss of signal.

However, for the (F) 255 K test, the recorded optical signal is 0.47 dB, with a maximum RIL of -2.57 dB – this signal is sufficient for data transmission (Figure 16B). There are two noticeable differences here from previous tests: (I) the test was performed after recalibration of the optical power meter, and (II) the test was performed at warm, ductile ice shell conditions.

### 4.7.2  HS STFOC Tether Signal Loss

The optical signal was successfully collected for HS STFOC tethers for the temperature range of 120 K to 248 K, with individual measurements at (A) 248 K, (B) 237 K, (C) 220 K, (D) 199 K, (E) 150 K, and (F) 120 K, as illustrated in Figure 18. Insertion loss relative to the first test reading was calculated to determine changes in signal due to velocity driving program (Table 3). There is an observed insertion loss of ~ -30 dB or -65 dB prior to freezing of tether in the ice or the start of shear testing, respectively. Three factors are potentially at play: (1) the unexplained high baseline loss (tare) of -30 dB, (2) alignment issues in fusion splicing, and/or (3) damage to the tether during install or pre-freezing in mold, which is harder to handle.

To resolve (1) ~ -30 dB tare in optical power meter–pigtail connection, the power meter was recalibrated by the manufacturer (OptoTest). Subsequently, the optical baseline was recorded at an optimal 1-2 dB (at 150 K). The validity of all prior results was then demonstrated by measuring the response (or, sensitivity) and relative signal dB loss (RIL) for the tether before and after recalibration, using "Pencil Wrap Tests" (Figure 17). Here, the patch cords were wrapped around a pencil and the dB versus number of wraps (up to 7) was logged to gauge sensitivity. The loss per wrap was constant, and the signal returned to the original baseline after unraveling the wraps (from 7 to 1), with almost identical increase in signal (~4 dB per wrap). This behavior is consistent before and after recalibration (Figure 17B, C).

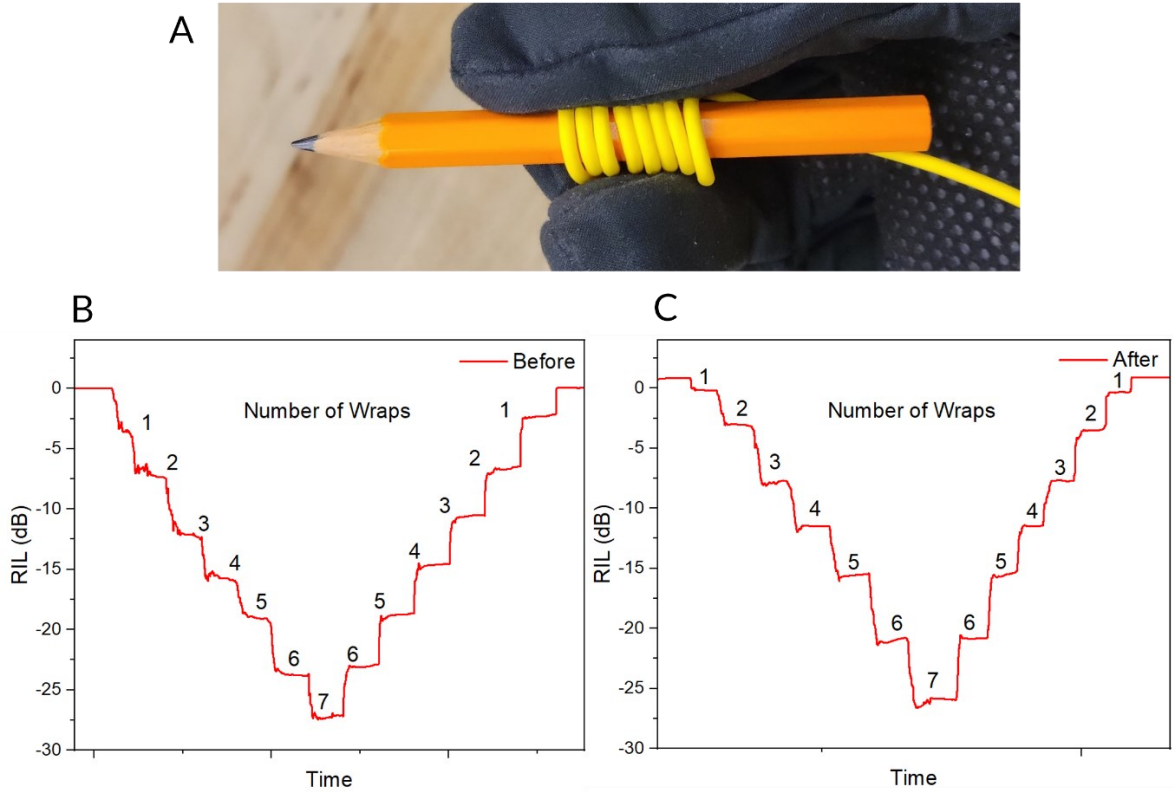

**Figure 17:** (A) Pencil wrap tests for HS STFOC tether with OP735 optical power meter: (B) before and (C) after recalibration by manufacturer to address a -30 dB tare: relative dB loss is accurately replicated indicating prior measurements are accurate.

Furthermore, during shear testing, the relative insertion loss in signal (RIL) is negligible at 248 K, 237 K and 220 K, with variations in the range of ±0.05 (min) to ±1.0 dB (max) observed during velocity ramps (creeping faults to ice quakes, respectively). This is similar to 255 K test results of the STFOC tether, where the warm, ductile ice surrounding the tether is deformed (and melted), thus accommodating some of the applied strain on the tether. At 248 K, an additional event with -5.9 dB drop is observed at 10:40 (Figure 18 A), which is assigned to human intervention (the tether was accidentally displaced). At ~11:05, a small amount (1.4 dB) of permanent damage is recorded, while at ~11:10, the meter again notes some strain but recovers quickly. For the 220 K test, a similar drop and recovery of -1.5 dB is recorded at ~13:05. Correlation between these signal drops and strain due to velocity ramps reveals the capabilities of tethers as strain gauges (Figure 18 A, C).

The low signal to noise ratios at 150 K and 199 K adds a ±6 dB error to the signal (RIL), which remains consistent throughout testing, and is difficult to correlate to velocity drops (Table 3). At ~150 K, there is an insertion loss of ~65 dB (from -1.6 dB during installation to -66 dB after freezing of ice-tether sample) during the ice manufacturing and freeze-in process. The absence of such a loss in previous tests implies issues due to (factor 3) crimping or bending of tether during manufacture or loading in this test. Further investigation (with spatial distribution of signal) is again required to identify the cause of signal loss prior to testing.

At the end of testing for 248 K, 199 K and 120 K, the HS STFOC tether was manually cut which coincided with a drop in signal to ~ -90 dB (purple curve), implying total loss. We can thus conclude that the tether was transmitting signal throughout testing, up to the time of failure due to manual cutting of the tether. Coupled with the consistency in optical signals recorded during testing at cryogenic temperatures, it increases our confidence in the data transmission abilities of HS STFOC tether for future missions.

### 4.8 Temperature Characterization

Both STFOC and HS STFOC cables were spliced with connectors, and pre-stress, ambient temperature measurements followed by non-ambient temperature (cryogenic) measurements were performed. The cryogenic temperatures were achieved using two methods: (i) tethers were placed (or submerged) for 72 hours in chest freezers at 253 K, 220 K, and a dewar filled with LN at 77 K. (ii) Tethers were placed in contact with a LN cooled cryostat and experienced a temperature range of 100 K to 273 K, with continuous measurements to capture evolution of optical signal during cooling and heating phases of testing.

In both tests, no significant losses in optical signal were detected, as highlighted in Figure 19A and B, with insertion losses recorded within 0 to -1.75 dB. The combination of test conditions in (i) and (ii), i.e., a temperature range of 77-273 K and duration of ~ 3 days, covers the range of conditions experienced by the two cables during shear (and optical) testing, but without any strain. This allows for a test of purely the impact of environmental conditions on the cables, which the two cables demonstrably survive in. Follow on testing with long duration cold soaks at 100 K are currently underway to capture the temporal aspects of tether survivability, and will be discussed in future work.

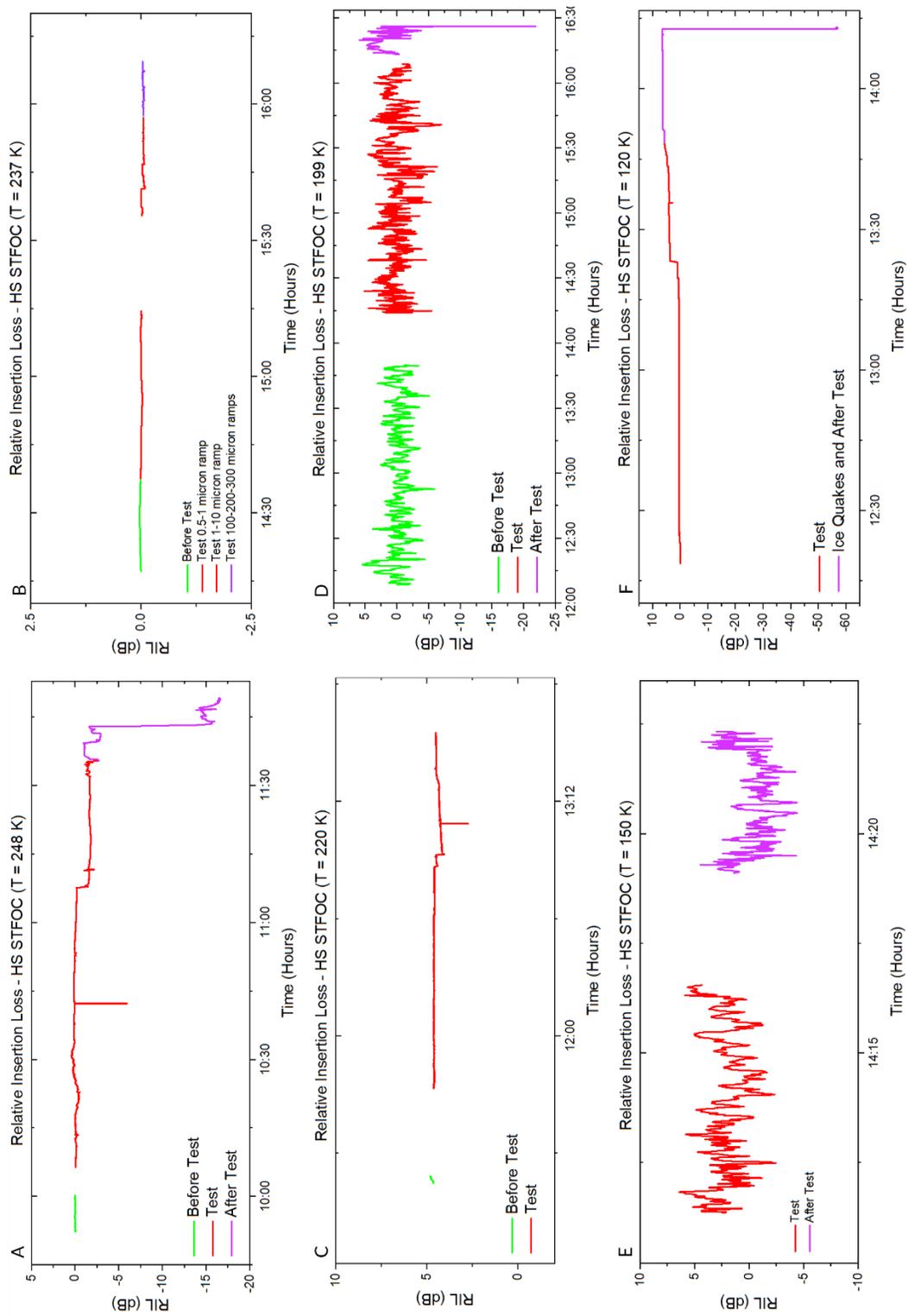

**Figure 18:** Relative Insertion Loss recorded for HS STFOC tether at: (A) ~ 250 K, (B) 237 K, (C) 220 K, (D) 199 K, (E) 150 K, and (F) 120 K, highlighting variations in signal (±0.5 to ~ ±7.0 dB) during velocity ramps.

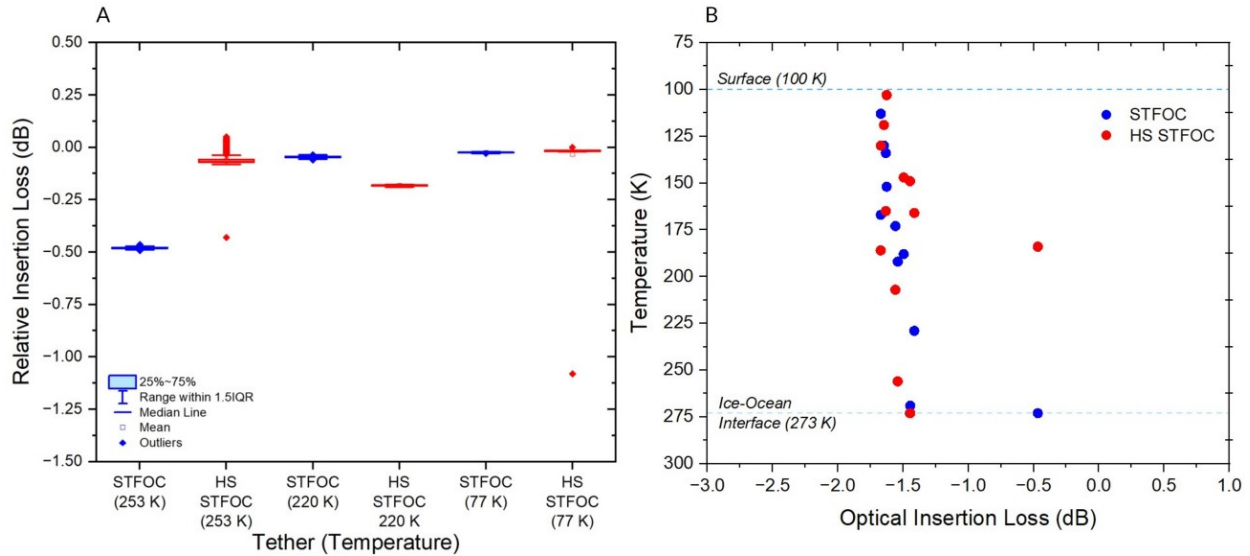

Figure 19: (A) Relative insertion loss recorded for STFOC and HS STFOC cables at 253 K, 220 K, and 77 K over three days, and (B) insertion loss over the temperature range of 100-273 K, highlighting the negligible drop in signal (< 1.75 dB) with temperature and time.

## 5. Discussion

*5.1 Do Tethers have a viable application to exploration of ice shells on Ocean Worlds?*

### 5.1.1 Laboratory Testing and Scientific Applications

Our cryogenic laboratory tests indicate that both the STFOC and HS STFOC tethers can survive displacements of 1.5 – 2.4 cm (current limit of the laboratory apparatus) across two fractures for a total tether length of 11.55 cm embedded in the ice (strain range of 0.13 – 0.42 m/m), while the ice experiences peak stresses of the order of ~1.3 to 1.5 MPa during initial loading. At the coldest temperatures tested (~100 to 120 K), subsequent high velocity events did contribute damage to the outer jacket, but that did not preclude the continuing functioning of the optical fiber. Comparing the peak stresses recorded for samples with and without tethers (Figure 8), another preliminary conclusion can be made that the presence of a tether might add resistance (or stress) to the surrounding ice, and potentially increase the breaking strength of the ice-tether assemblage leading to cracking at the two ice-ice interfaces, at a higher peak stress.

From a communication standpoint, (1) in STFOC tether tests, we were unable to record a high baseline signal suitable for communication, post sample fabrication, apart from measurement at a relatively warm temperature of 255 K. (2) However, in HS STFOC tether tests, the optical meter readings confirm that tethers have a higher baseline signal with some insertion loss observed during fabrication, and do not display significant loss during the ice fracture process. Evidently, the presence of additional layers of a Kevlar strength member and a TPU jacket increases the survivability of tethers. Thus, we can state that HS STFOC tethers can potentially survive conditions akin to reactivation of faults at cryogenic temperatures on Ocean Worlds, and may offer a viable pathway for exploration of Europa's interior, although further tether design development would have the potential to increase robustness relative to the more extreme fault conditions that could be anticipated.

Furthermore, the distributed sensing capabilities of tethers are extremely exciting for Ocean World exploration. The ability to use tethers as strain gauges to record strain profile of the ice shell, estimate the temperature profile, and constrain fault orientations and motions at depth provides additional mission capabilities for investigating properties *within* the interior of the ice shell, in addition to providing a conduit for communications *through* the ice.

### 5.1.2 Limitations

Although the tethers survived the shear tests, at the coldest temperatures (~100 K) the outer jackets were damaged, and inner fibers stretched, helping guide priorities for future tether development. Further, the shear tests performed do not represent the full range of potential damage that may be inflicted on a tether, both in terms of geometry of stressing and chemistry of surrounding ice. For example, the Galileo Near-Infrared Mapping Spectrometer (NIMS) data revealed predominant non-ice components on the heavily radiation-bombarded trailing hemisphere of Europa, including spectra that could be interpreted as $MgSO_4$ hydrates, sulfuric acid hydrates, and/or sulfate salts (Carlson et al. 2002, 2009; McCord et al. 1998). Subsequent work has shown that additional salts, including NaCl (Trumbo et al. 2019; Hand & Carlson 2015), Mg-bearing chlorinated species (Ligier et al. 2016; Johnson et al. 2019) could also be present, along with and Na and Mg sulfates (Fischer et al. 2015; Vu et al. 2016; Johnson et al. 2019). Some of these non-water-ice constituents appear to be particularly concentrated in association with large-scale geologic units such as chaos and regions containing dark colored materials (McCord et al. 1998). Other Ocean Worlds are suggested to have ammonia in the ice shell, such as Enceladus (Squyres et al. 1983; Kargel 1992; Kargel & Pozio 1996; Waite et al. 2009), Titan (Fortes et al. 2007), and Triton (Hogenboom 1997). Optical tethers have never been tested for survival within these chemistries, which might potentially damage the outer jacket or lead to optical propagation loss due to reaction with the strength members in the tether.

Optical transmission results indicate a need to eliminate signal loss prior to testing, with improved fusion splicing, handling, and/or loading of tethers in the sample. The recorded test signal for STFOC tethers is currently insufficient for successful data transmission and could potentially rule them out for application in any future mission. However, before such a classification is made, a new set of spatial measurements are required with a high-resolution Optical Backscatter Reflectometer (OBR), to investigate the source of insertion loss. Using a Luna OBR 4600, we were able to perform a proof of concept for optical transmission and distributed sensing during shear testing: to monitor changes along the 3-meter-long fiber, at a spatial resolution of 10 microns. Our results demonstrate that this unit was capable of identifying the location of (i) insertion loss (IL), and (ii) shear planes (or faults) between central and side blocks.

Also, ongoing modeling efforts by our team (e.g., Lien et al. 2022; in prep) have identified the need for further shear testing at greater displacements, which can be achieved by modifying the driving velocity program in future work.

### 5.2 Sliding Behavior of the Ice Shell

The identical sliding behavior with velocity and temperature, as described for tether samples and control ice samples, demonstrates two key points:

1. In these shear experiments, the tether is not radically affecting the ice response; instead, the tether appears to be "along for the ride".
2. Our study with the crudely broken and imperfectly planar sliding interfaces, similar to reactivated faults on Europa, confirms the previously well-constrained results of frictional sliding behavior of Schulson & Fortt (2012). It provides robust confirmation of the distinct temperature and velocity dependence of frictional sliding in ice, which is significant for its potential application on Europa and other icy worlds.

Further, if the stability of friction is so closely tied to temperature, as demonstrated here, it suggests that a "seismogenic zone" may be present at depth on Europa, as has been identified on Earth (thought instead to be due to creation of rock gouge with depth, Marone and Scholz, 1988). Although it is beyond the scope of this work to map a zone of instability with depth on Europa, continued work in that area is being conducted by collaborators (McCarthy et al. 2018 & 2022; Zaman et al. 2021). The results presented here (consistent with Schulson and Fortt, 2012) suggest that there may be a variation of sliding behavior with depth such that the uppermost and lowest portions of the ice shell would be sliding smoothly (and slowly), whereas at a mid-range in temperature and depth, icy faults could initiate stick-slip, rapid Europa-quake events. Thus, Europa may naturally be providing a wide range in temperature and sliding velocity conditions that are fairly consistent with the conditions that we have tested our tethers against here.

Our laboratory results and Brittle to Ductile Transition (BDT) temperature cannot be directly scaled to Europa because BDT is not only temperature dependent, but it also depends on confining pressure and on strain rate (a broad range of which is captured in laboratory testing). On Europa, calculations of BDT temperature and depth require constraints on thermal gradient and ice chemistry. For example, with the addition of salts to the ice shell (e.g., two phase ice-$MgSO_4$ eutectic composition), the strength envelope (and BDT transition) may include a zone of semi-brittle behavior (McCarthy et al. 2011). Future tests will evaluate the role of chemistry in the observed phenomenological BDT transition.

### 5.3 Future Work

Our work represents an important contribution to enabling the future exploration of an Ocean World, including the search for extraterrestrial life, through evaluation of tethered techniques as a means of communication with a cryobot, tunneling its way through an ice shell to access the underlying ocean. We have investigated a tethered architecture that could survive and operate effectively on Jupiter's moon, Europa, which has high potential for a near term subsurface exploration missions (Howell et al. 2020). We have attempted to raise the Technology Readiness

Level (or TRL) of tethered communication technology to 4, through a robust set of laboratory tests, thereby eliminating some of the technical risk to the overall penetration system. The environmental conditions simulated in our laboratory tests map to a range of predicted changes in conditions at different depths for Europa's ice shell (e.g., temperatures, ice mechanical behavior, and sliding fault velocities). However, these conditions and test results are also relevant to other icy worlds in the outer solar system.

On a subsystem scale, we have identified the limitations of commercially available tethers, and identified the Linden Photonics Inc. High Strength STFOC tether (HS STFOC) as a viable selection for future testing. Our team will further design, build, and test robust optical communication tether prototypes for Ocean World conditions. From a laboratory perspective, our team plans to explore additional material properties (e.g., adhesive strength) and performance of the additional tethers with different jacketing material and strength layers, focusing on ideal performance at specific temperature/chemical conditions and considerations for reducing overall mass. Results from these tests can be used to develop a custom prototype tether with enhanced strain tolerance that incorporates a "loose tube" construction, which in turn will be performance tested. We will also focus on strain and temperature measurements using the Luna OBR 4600 instrument to create strain and thermal profiles of ice samples during laboratory testing.

From a modeling perspective, the cycle of tidal bulging on Europa is expected to result in a back-and-forth motion along the fault, but fault displacements are poorly constrained. There is a need to globally explore faults on Europa's surface and over the entire tidal cycle to simulate when fault "events" would be likely to occur, and at what depth, to determine overall total deformation.

The next phase of research beyond that outlined here would be to implement a program of representative testing in the laboratory coupled to a model cryobot, using a combination of cryogenic chambers to reproduce the extremely cold, hard and brittle conditions expected at the outer shell of Europa or Enceladus. Subsequent field deployments drilling into Earth's cryosphere will approach the warmer and more ductile conditions to be encountered deeper, closer to and at ice-water interfaces, where conditions particularly conducive to life are anticipated. As more data, and better constraints, become available (e.g., from the Europa Clipper mission), we will be well-positioned to assess the efficacy of available tethers and identify any required improvements to make them viable within Europa's ice shell.

Finally, there is a need for continued investment in subsurface access technology development and Ocean World modeling investigations to improve our current understanding of ice shell and ocean properties (to identify environmental hazards), and to develop sub-systems to a maturity ready for preliminary design within the next decade.

## Acknowledgements

The authors would like to acknowledge funding from the NASA Scientific Exploration Subsurface Access Mechanism for Europa (SESAME) (80NSSC19K0613), and COLDTech: Autonomy, Communications, and Radiation-Hard Devices (80NSSC21K0995) and technology development opportunities at Applied Physics Laboratory, Johns Hopkins University for this work.

VS acknowledges Diana Catalina Sanchez Roa for her contributions to SEM imaging and microscopic analysis of STFOC tethers. The authors acknowledge the two anonymous reviewers for their valuable insight which helped improve and clarify the presentation of this article.

**Appendix A: Predicted Stress, Strain, and Shear Velocity**

In lieu of observational evidence to provide event slip rates on Europa, we use basic relations from elasticity that state that the ratio of propagation speed to slip speed nearly equals the ratio of the elastic shear modulus to peak-to-residual stress drop at the front (Rubin 2011). The average stress drop $\Delta\tau$ can be estimated from elastic properties (effective shear modulus $\mu'$) and an assumption about the downdip extent of slip W (Pollard and Segall 1987), using equation A1:

$$\Delta\tau/\mu' = \delta/W \qquad \ldots \text{(Equation A1)}$$

From kinematics, the maximum slip speed, $V_{max}$, equals the propagation speed, $V_{prop}$, multiplied by the maximum slip gradient. Combining these two relationships provides equation A2:

$$V_{prop}/V_{max} = \alpha\,(\mu'/\Delta\tau_{p\text{-}r}) \qquad \ldots \text{(Equation A2)}$$

where $\alpha$ is a scaling factor that depends on the spatial distribution of strength reduction behind the peak stress, but is close to 1 for quasi-static elasticity and subscript p-r stands for peak-to-residual (e.g., Ampuero & Rubin 2008). As an estimate for propagation speed, we take the shear wave speed velocity through ice (~2000 km/s). Based on Rubin 2011 slip relationships and the parameter values for ice, slip rates during an event may be as fast as $10^{-3}$ m/s. Consequently, we explored velocities in the range $10^{-7}$ to $10^{-3}$ m/s, to simulate these end members, and determined their survivability in ice.

**Appendix B:** Classic Examples of Frictional Behavior of Ice: Shear Stress Dependence on Velocity

Laboratory friction experiments have long been used to determine rock and ice frictional strength and stability as a way of understanding the mechanics of earthquakes and faulting (e.g., Dieterich 1972; 1978; 1979; Scholz 1998, 2019) and more recently, glacier sliding (e.g., Zoet et al. 2013; McCarthy et al. 2017). It is beyond the scope of this work to go into a detailed exploration of friction and faulting. However, since the experiments performed are an adaptation of a double-direct-shear friction experiment, it is worth describing some of the classic and representative features of experimental friction data. Figure B1 shows a typical response from a velocity step and a slide-hold-slide. The data is from a system of ice sliding on rock, but nearly identical responses are seen for rock samples (e.g., Dieterich 1972; Beeler et al. 1994; Marone 1998).

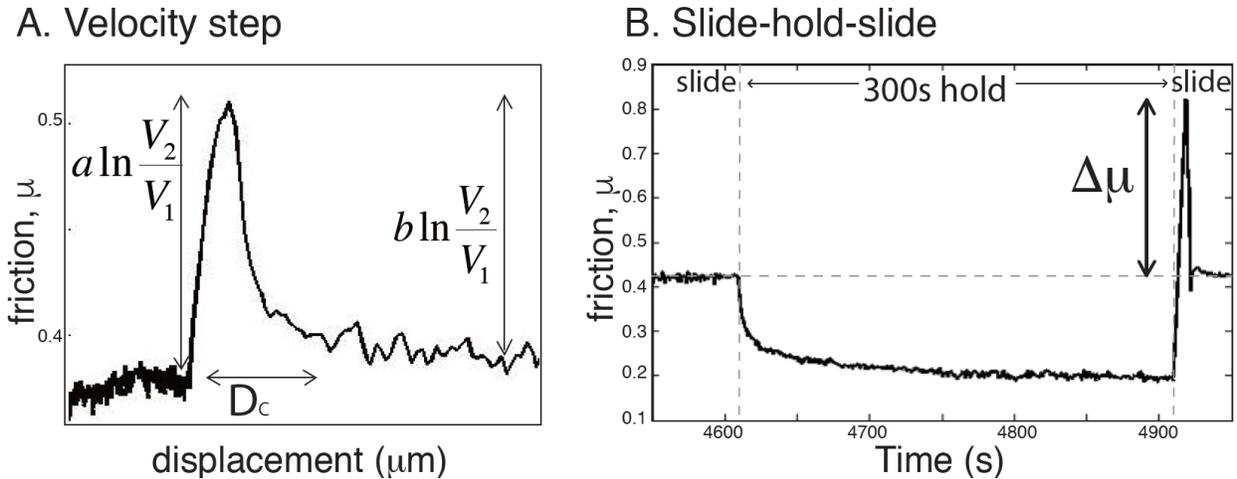

**Figure B1:** Example of (A) single velocity step used to determine rate-state friction parameters (a, b), and (B) a slide-hold-slide which consists of relaxation of friction on the onset of hold, and a peak when sliding is resumed. Here, $D_C$ is the critical slip distance needed for system to evolve from one steady-state friction to another, parameters μ, a and b are experimentally determined coefficients, and $V_1$, $V_2$ are sliding velocities. The examples are from polycrystalline ice sliding between two granite blocks (McCarthy et al. 2017).

In the example, an order of magnitude step up in velocity was employed. The frictional response is an immediate increase in friction (the "direct effect") and then friction evolves back down to a new steady state friction value. Velocity steps are analyzed to determine the so-called rate- and state-dependent friction parameters (e.g., Dieterich 1978; Ruina 1983). Not shown here, but in systems of granular interfaces like till or gouge, the direct effect is not so dynamic and peak-like. Instead, friction in granular systems demonstrates a gradual hill shape with a velocity increase (e.g., Chester 1994).

During slide-hold-slides, the driving piston moves forward at a constant rate and then is held constant for increasing durations (in Figure B1 B, 300 s). The piston does not move backward, so the friction does not fall to zero. Rather, the held piston causes the friction to slowly relax. The relaxation is due to a combination of the apparatus itself and the sliding interface. Upon sliding again, the friction jumps up due to time-dependent healing at the sliding interface. That change in friction after a hold typically increases according to the log of time (e.g., Beeler et al. 1994). Healing is generally attributed to increase of real contact area, which consists of the asperities that are in contact at the microscopic level. In ice, healing has been known to be quite large compared to that of rock due to the high homologous temperature and more easily deformable asperities (e.g., McCarthy et al. 2017; Zoet & Iverson 2018).